\documentclass[12pt]{article}
\pdfoutput=1
\usepackage{wrapfig}
\usepackage{epsfig}
\begin{document}

\begin{center}

\vspace{0.5cm}

{\bf \Large The Higgs Boson is found: What is next?}
\renewcommand{\thefootnote}{\fnsymbol{footnote}}\footnote[2]{Talk given at a scientific session of the Department of Physical Sciences of the Russian Academy of Sciences, February 26,  2014}
\vspace{1cm}

{\bf \large  D.I.Kazakov}\vspace{0.5cm}

{\it Bogoliubov Laboratory of Theoretical Physics,\\
Joint Institute for Nuclear Research, Dubna\\[0.2cm]
Alikhanov Institute for Theoretical and Experimental Physics, Moscow\\[0.2cm]
Moscow Institute of Physics and Technology, Dolgoprudny}\vspace{0.6cm}

\abstract{The situation in particle physics after the discovery of the Higgs boson is discussed. Is the Standard Model complete? Are there still mysteries which have no answer? Answering these questions we consider the Higgs sector, the neutrino sector and the flavor sector of the Standard Model, and list the problems which are still far from understanding. Going beyond the Standard Model we consider the Dark matter in the Universe and possible existence of new particles and interactions. The main attention is paid to supersymmetry. The problems faced by elementary particle physics in the near perspective are formulated.
}
\end{center}

\section{Introduction} 
\begin{wrapfigure}{l}{0.5\linewidth} 
\begin{center}\vspace{-0.7cm}
\leavevmode
\includegraphics[width=0.5\textwidth]{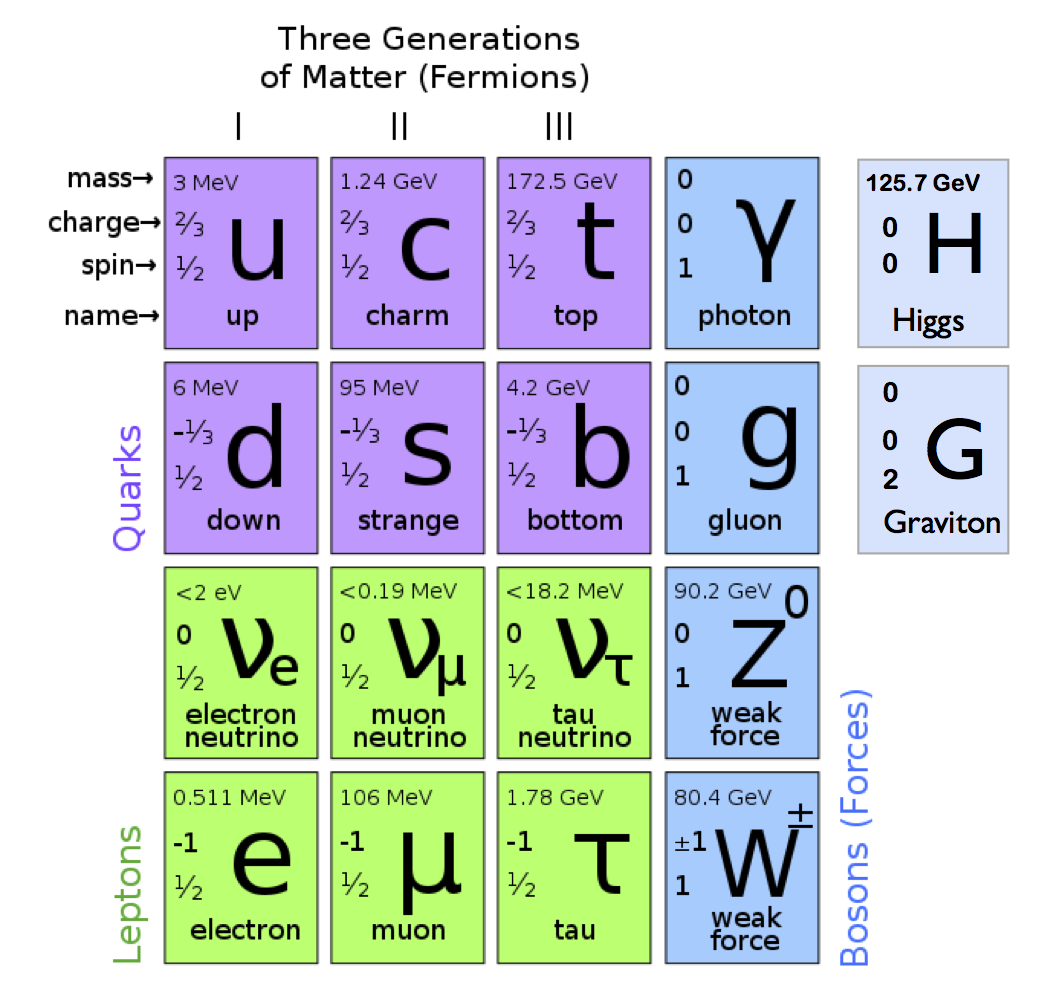}
\end{center}
\caption{The Standard Model of fundamental interactions~\cite{SMPicture}}
\label{SM}
\end{wrapfigure}

The discovery of the Higgs boson in 2012~\cite{HiggsDis1,HiggsDis2} and the Nobel prize award in 2013 
have marked an important step in elementary particle physics. The mechanism of fundamental particle mass generation, the Brout-Englert-Higgs mechanism~\cite{BE,H},  theoretically predic\-ted nearly 50 years ago, is experimentally confirmed. Thus, the Standard Model of fundamental interactions has got its logical completion and obtained the status of the Standard Theory. By the Standard Model we understand the description of strong, weak and electromagnetic interactions between quarks and leptons based on the  gauge group  $SU(3)_c\times SU(2)_L\times U(1)_Y$. Here quarks are triplets and leptons are singlets
with respect to the color group $SU(3)_c$, the left components of quarks and leptons are doublets and the right components are singlets with respect to $SU(2)_L$, and all of them possess a hypercharge, according to the group $U(1)_Y$. The set of matter fields and the carriers of four fundamental forces
of the SM are shown in Fig.\ref{SM}.  To the already known particles, all of which being discovered in the XX century, one should add the Higgs boson discovered  in the XXI century.

In the SM one has six quarks and six leptons forming three generations and three types of interactions: strong, weak and electromagnetic mediated by the quanta of the corresponding fields:  the gluon, the W and Z bosons and the photon. One should add now the fourth interaction, the Yukawa one carried  by the Higgs boson. For completion of the picture one should add also the gravitational interaction mediated by the quantum of  the gravitation field, the graviton. However, it is not discovered yet: gravity remains still a classical theory. This exhausts all the known fundamental particles and forces in Nature.
 \phantom{\hspace{5cm}}
The Standard Model of strong, weak and electromagnetic interactions quantitatively describes practically all experimental data. There is no experiment in particle physics where  the deviation from the SM exceeds 2-2.5 standard deviations and sometimes appearing  anomalies disappear with time~\cite{SMtest}. It might be that the recently discovered neutrino oscillations will require some minor modification of the SM. However, this is not necessary: Addition of the right handed neutrinos is fully inscribed in the SM and allows one to describe the transformation of one kind of neutrino into another due to the mixing, as it happens in the quark sector. New precision tests of the processes with flavor changing and with CP-violation have also passed all  checks. Remarkably, all experiments are described with the help of a single Cabibbo-Kobayashi-Maskawa matrix with three parameters. Thus, we are witnessing the triumph of the Standard Model of fundamental interactions as the basis of all phenomena in Nature, except gravity.

The natural question arises: Is this the end of the story and its new stage? The answer given by the scientific community does not leave any doubt: This is the beginning of a new research program for a few decades. Nature still keeps many puzzles! 

\begin{figure}[ht]
\begin{center}
\leavevmode
\includegraphics[width=0.8\textwidth]{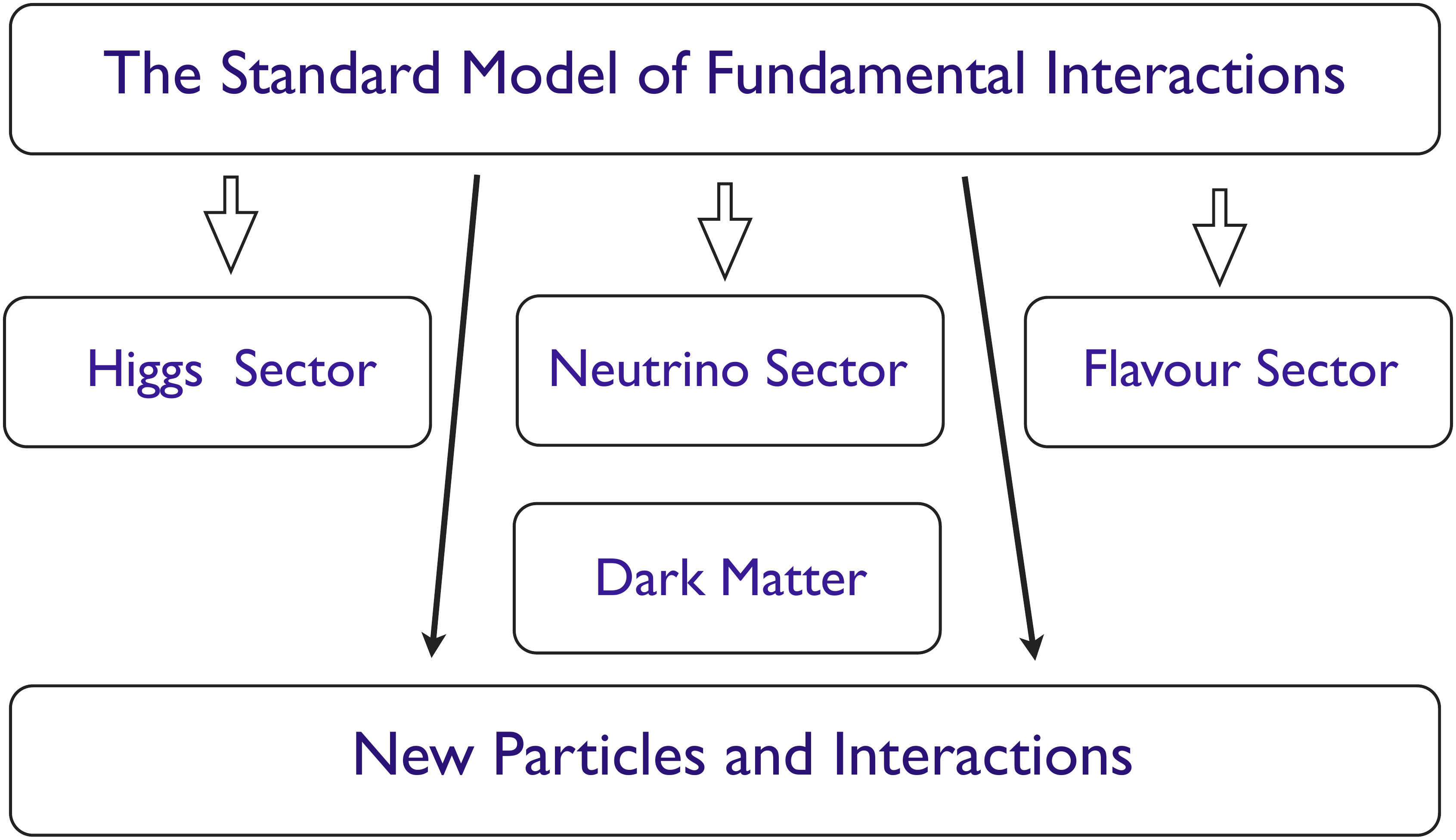}
\end{center}
\caption{The problematic sectors of the Standard Model and beyond}
\label{diag}
\end{figure}

Discussing various aspects of the Standard Model and some attempts to go beyond it we follow the schematic diagram shown in Fig.\ref{diag} and consider all its elements in detail.

\section{The Higgs sector}

Thus, the Higgs boson is discovered. At the  99\% confidence level its properties are determined and they are in  good agreement with the expectations: This is a particle with spin 0, parity +, nonzero vacuum expectation value, it interacts with W and Z bosons as well as with quarks and lepton (checked for the third generation) with the strength proportional to their masses~\cite{HiggsCouplings}. Still, the exploration of the Higgs sector of the SM just begins. The formulated questions require answers:
\begin{itemize}
\item Is it the Higgs boson?     -    Most probably, yes.
\item Is it  the Higgs boson from the Standard Model?  - It looks like.
\item Are there alternatives?  -  Yes.
\item Can it be that we see more than one Higgs boson?  - Possibly.
\item It is possible  to get the reliable answers to these questions?  - Yes.
\end{itemize}
The new experiments at the LHC at the doubled energy and at new accelerators (if built) will allow one to reach the required accuracy for unambiguous answers to the above questions. Note, however, that we have already got the confirmation that particles get their masses through the Brout-Englert-Higgs mechanism no matter what model the Higgs boson corresponds to. 
 
\begin{figure}[ht]\vspace{0.6cm}
%\begin{center}
\leavevmode \hspace{1cm}
\includegraphics[width=0.40\textwidth, height=7cm]{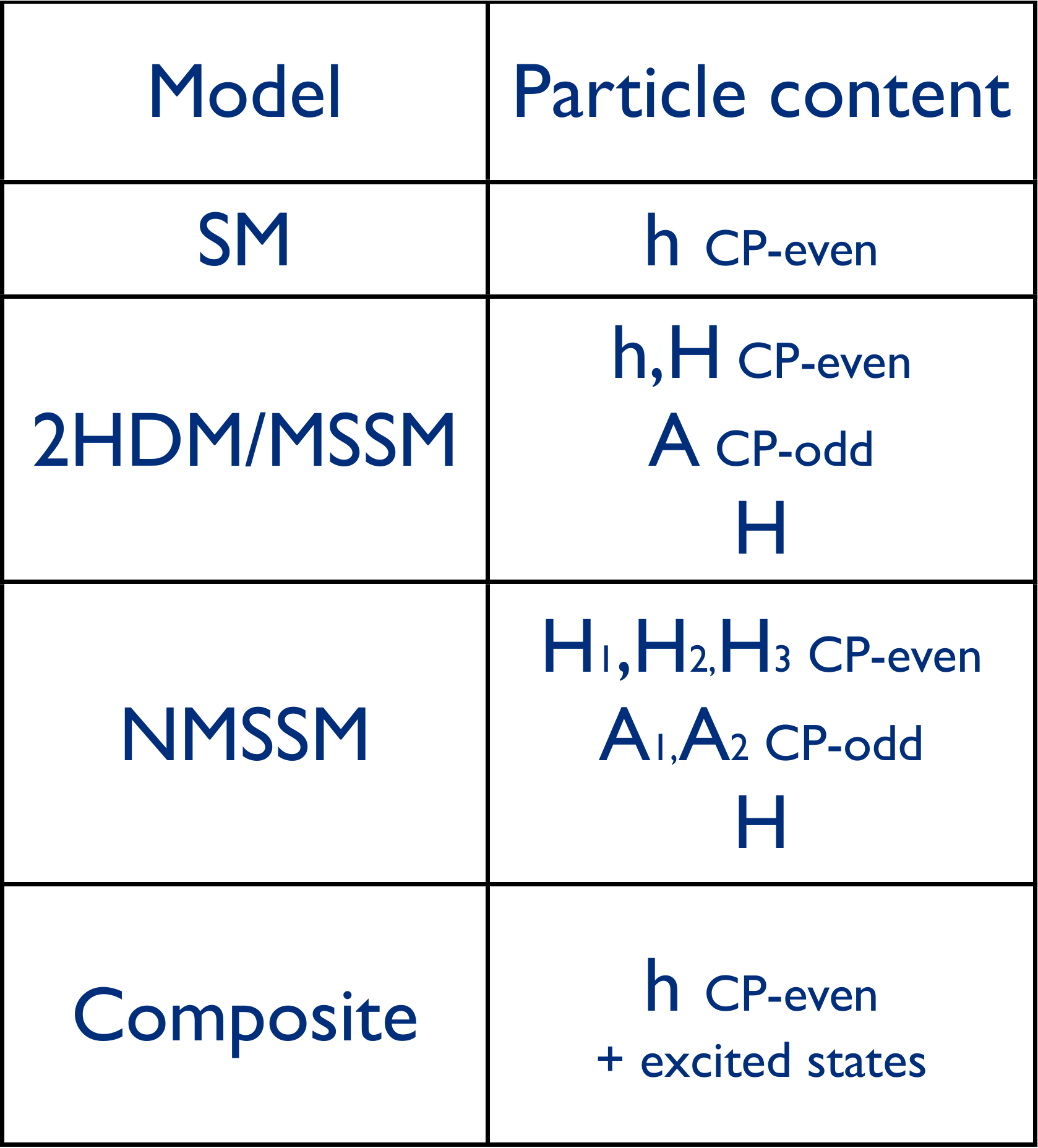} \vspace{-7cm}

\hspace{8.4cm}\includegraphics[width=0.4\textwidth,height=7.4cm]{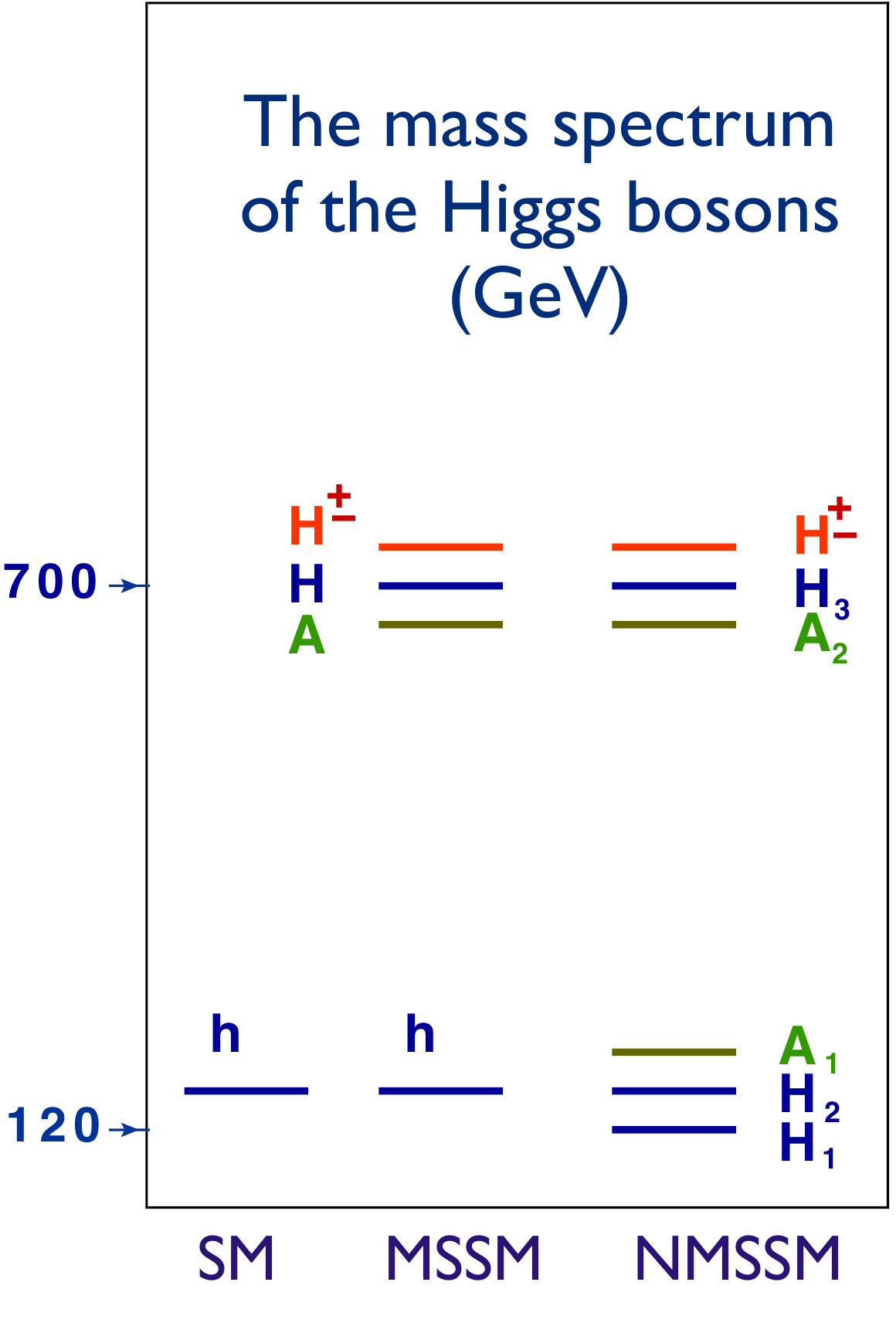}
%\end{center}
\caption{The field content and the spectrum in various models of the Higgs sector}
\label{spectr}
\end{figure}

Consider possible alternatives to the minimal Higgs sector. Remind that the minimal  SM contains one Higgs doublet which provides  up and down quarks and leptons with  the mass simultaneously. In this case, there is only one CP-even Higgs boson (see Fig.\ref{spectr}, left). The nearest extension of the SM is the two Higgs doublet model~\cite{THDM}. It is also realized in the case of the Minimal Supersymmetric Standard Model (MSSM)~\cite{MSSM}.  Here the up and down quarks and leptons interact with different doublets  each of which has a vacuum expectation value. In this case, one has 5 Higgs bosons:  two CP-even, one CP-odd and two charged ones. 

The next popular step is the introduction of an additional Higgs field which is a singlet with respect to the gauge group of the SM. In the case of supersymmetry, this model is called the NMSSM, the next-to minimal~\cite{NMSSM}.  Here one has already seven Higgs bosons. The sample spectrum of particles for various models is shown in Fig.\ref{spectr}, right. Note that in the case of the NMSSM one has two light CP-even Higgs  bosons and the discovered particle might correspond to both $H_1$ and to  $H_2$. The reason why we do not see the lightest Higgs boson $H_1$ in the second case is that it has a large admixture of the singlet state and hence very weakly interacts with the SM particles.

Finally, it is possible that the Higgs boson is the composite state like the  $\pi$-meson~\cite{composit}. Then, besides the ground state there should be the exited heaviar states.

In all these cases one of the Higgs bosons is very close by its properties to the SM Higgs boson and quite possible that we see just one of these states.  Thus, the Higgs sector needs to be explored. It is necessary to be convinced in the presence or absence of heavy and charged Higgs bosons.

\begin{wrapfigure}{l}{0.6\linewidth} 
\begin{center}\vspace{-0.2cm}
\leavevmode
\includegraphics[width=0.6\textwidth]{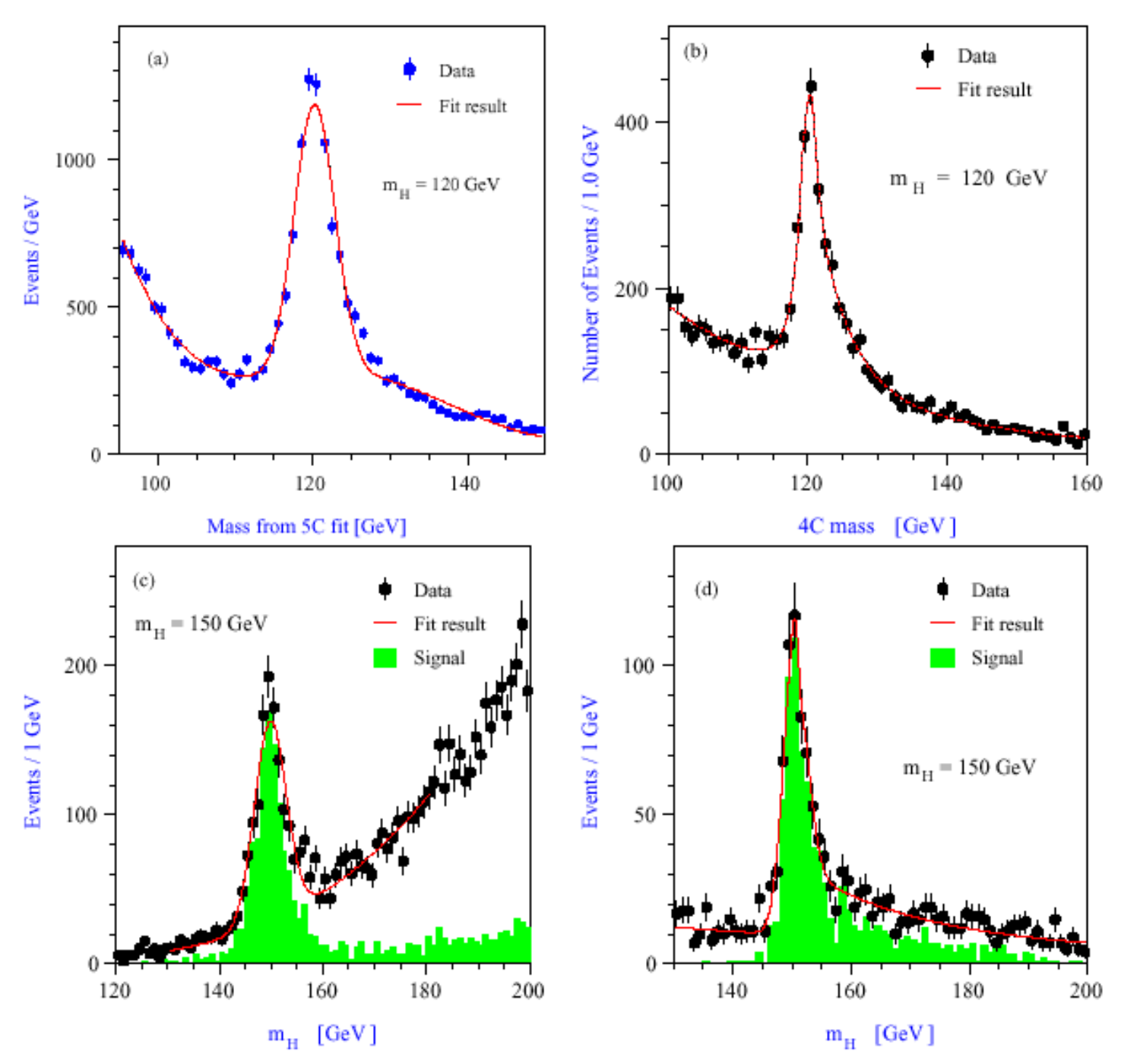}
\end{center}\vspace{-0.2cm}
\caption{The measurement of the mass and the width of the Higgs boson in various channels at the ILC:
$e^+e^-\to HZ\to b\bar b q\bar q,  q\bar q l^+l^-, W^+W^-q\bar q,  W^+W^-l^+l^-$ }
\label{ILC}
\end{wrapfigure}

\begin{figure}[ht]
\begin{center}
\leavevmode
\includegraphics[width=0.33\textwidth]{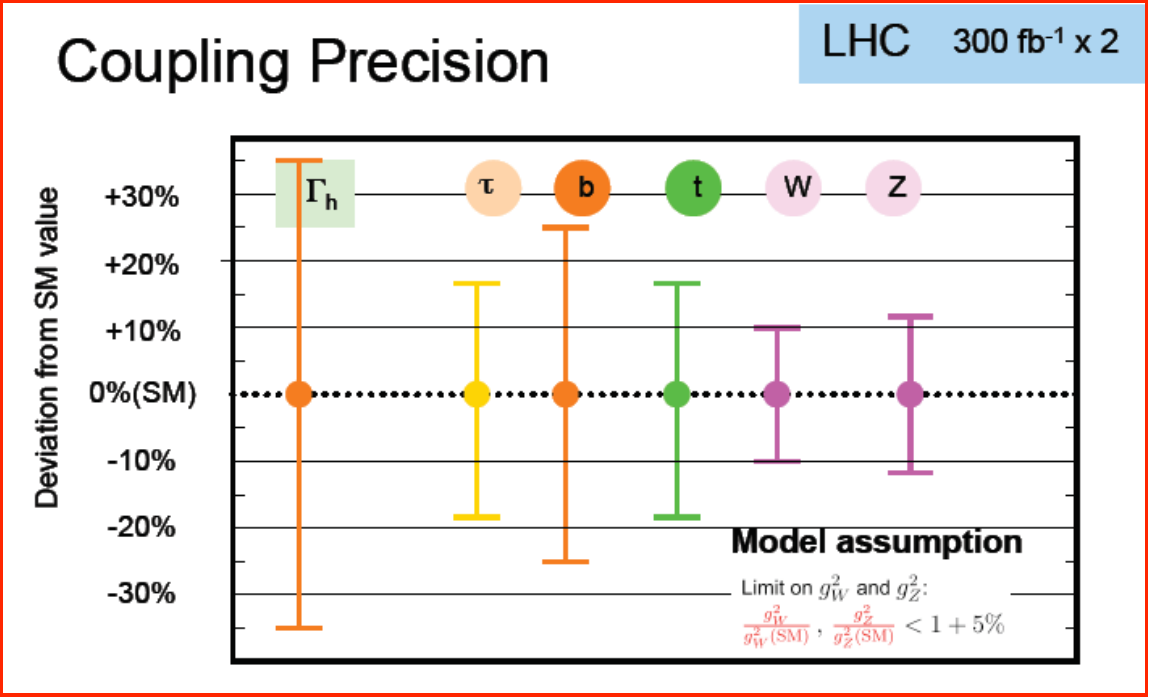}
\includegraphics[width=0.315\textwidth]{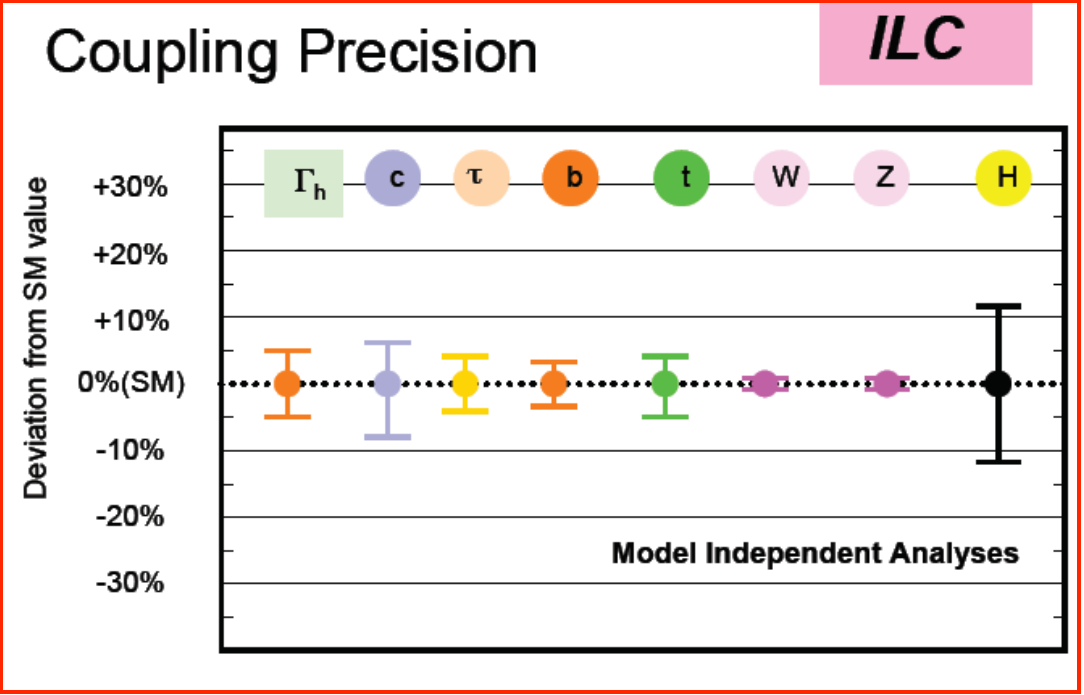}
\includegraphics[width=0.315\textwidth]{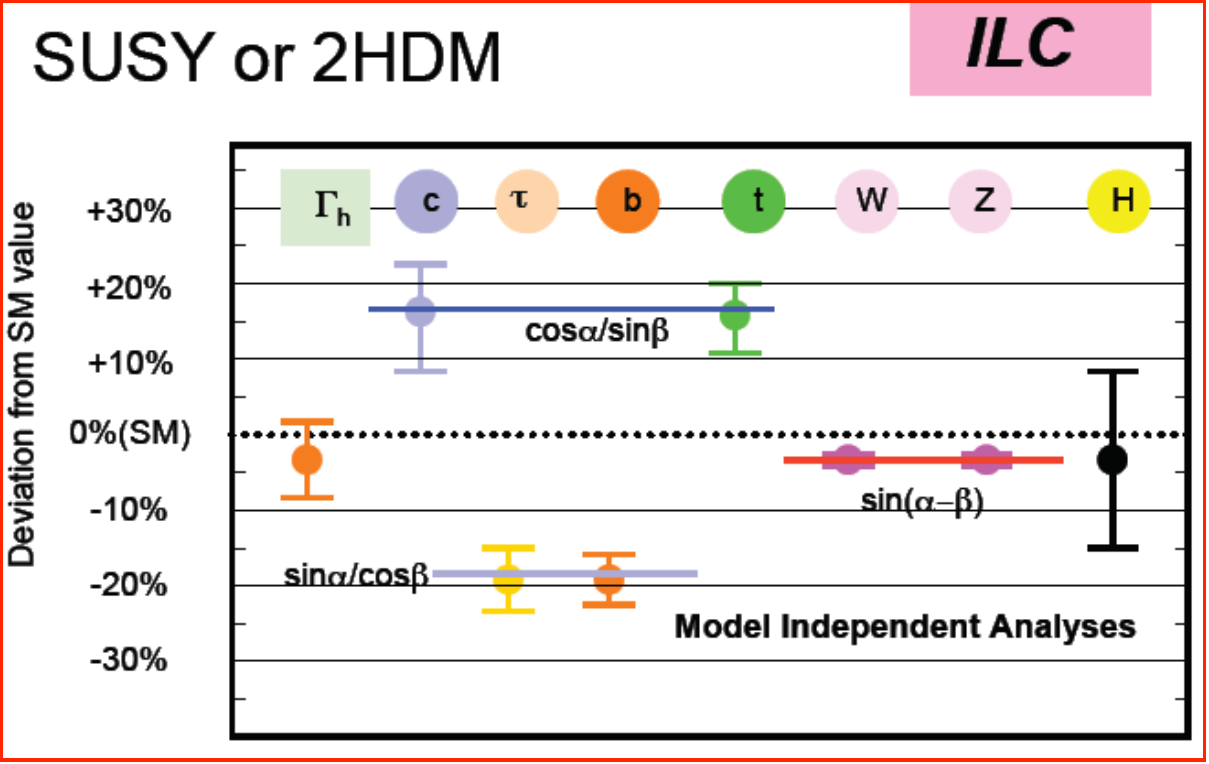}
\end{center}
\caption{The measurement of the Higgs boson couplings at the LHC and ILC~\cite{couplings}}
\label{couplings}
\end{figure}

The task for the near future is the precision analysis of the discovered Higgs boson. It is necessary to measure its characteristics  like the mass and the width and also all decay constants with the accuracy  ten times higher than the reached one. Quite possible that this task requires the construction of the electron-positron collider, for instance,  the linear collider ILC.  Figure \ref{ILC} shows the expected results for the Higgs boson mass measurement at the ILC in various channels~\cite{ILC}. \phantom{\hspace{4cm}}
It is planned that the accuracy of the Higgs mass measurement  will achieve $\sim$50  MeV that is 5-7 times higher than the achieved one.  Another task is the accurate determination of the constants of all decays which will possibly allow one to distinguish the one-doublet model from the two-doublet one. Figure \ref{couplings} shows the planned accuracies of the measurement of the couplings of the Higgs boson with the SM particles at the LHC for the integrated luminosity   of
300 1/fb (left), which is ten times higher than today.  For comparison we show also the same data for the ILC (middle). The accuracy of measurement of the couplings at the ILC will allow one  not only to distinguish different models but also check the predictions of supersymmetric theories (right).

\section{The neutrino sector}

With the discovery of neutrino oscillations neutrino physics has entered the new phase: The mass differences of different neutrino  types and the mixing angles were measured.  At last, the answer to the question of neutrino mass was obtained. Now we know that neutrinos are massive. This way, the lepton sector of the SM took the form identical to the  quark one and it was confirmed that the SM possesses the quark-lepton symmetry.  Nevertheless, the reason for such symmetry remains unclear, it might well be that it is a consequence of the Grand unification of interactions. However, the answer to this question lies beyond the SM. 

\begin{figure}[ht]
\begin{center}
\leavevmode
\includegraphics[width=0.8\textwidth]{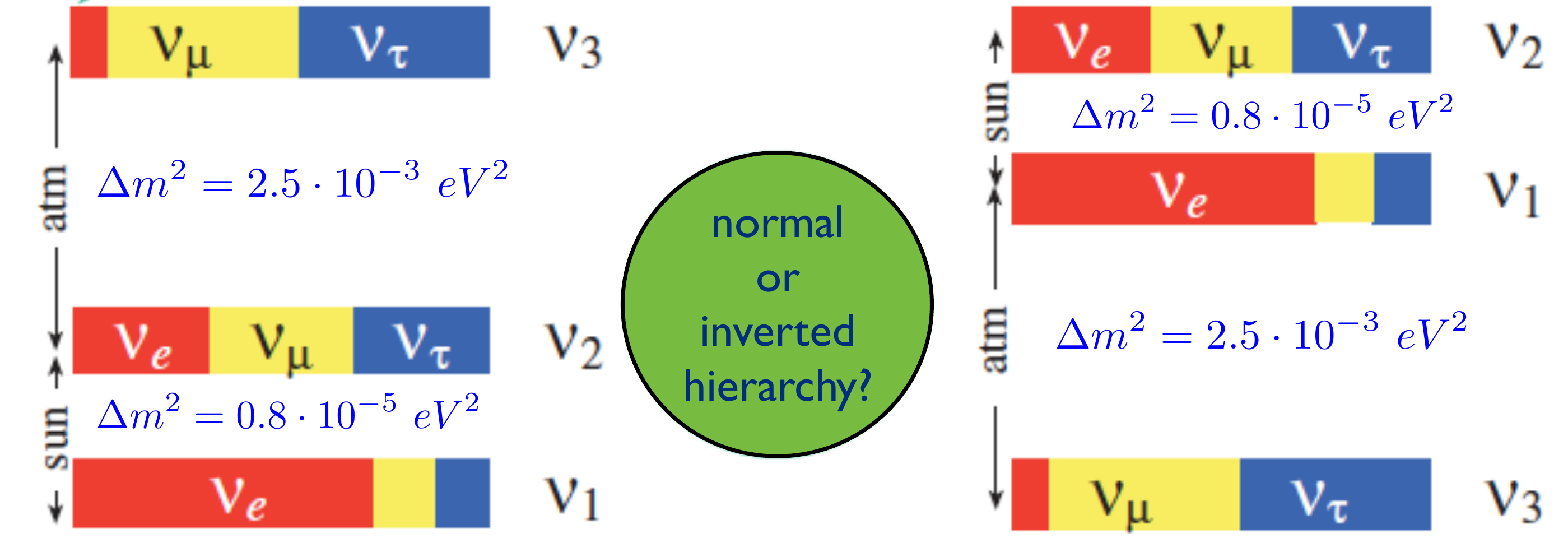}
\end{center}
\caption{Normal and inverse hierarchy of neutrino masses}
\label{neutrino}
\end{figure}

At the same time, the neutrino sector of the SM is still not fully understood.  First of all, this concerns the mass spectrum.  Neutrino oscillations allow one to determine only the squares of the mass difference for various neutrinos. The obtained picture is shown in Fig.\ref{neutrino}~\cite{neutrino}.  The color pattern shows the fraction of various types of neutrino in mass eigenstates. 

Besides the hierarchy problem (normal or inverted) there is also an unclear question of the absolute scale of neutrino masses.  One may hope to get the answer to this question in two ways. The first one is the direct measurement of the electron neutrino mass in the $\beta$-decay experiment. According to the Troitsk-Mainz experiment, the upper bound on the neutrino mass today is $m_{\nu_e}< 2$ eV ~\cite{TroitskMainz}. The preparing experiment KATRIN~\cite{Katrin} will be able to move this bound up to
$<0.2$ eV. However, this might not be enough if one believes in astrophysical data. The determination of the sum of neutrino masses from the spectrum of the cosmic microwave background  is an indirect but rather an accurate way to find the absolute mass scale. At the early stage of the Universe during the fast cooling
process particles  fell out of the thermodynamic  equilibrium  at the temperature proportional to their masses and their abundance ``froze down" influencing the spectrum. Hence, fitting the spectrum of the CMB fluctuations one can determine the number of neutrino species and the sum of their masses~\cite{CMBNeutrino}.  The result of the latest space mission PLANK~\cite{Planck} looks like $\sum m_\nu <0.23$ eV. This number is still much bigger than the neutrino mass difference shown in Fig.\ref{neutrino}.
Thus, the absolute scale of neutrino masses is still an open question.

Another unsolved problem of neutrino sector is the nature of neutrino: Is it a Majorana particle or a Dirac one, is it an antiparticle  to itself or not?  Remind that  particles with spin 1/2  are described by the Dirac equation, the solutions being the bispinors. They can be divided into two parts corresponding to the left or right polarization
\begin{equation}
\nu_D=\left(\begin{array}{c} \nu_L \\ 0 \end{array}\right)+\left(\begin{array}{c} 0 \\ \nu_R \end{array}\right), \ \ \ \nu_L \neq \nu_R^*, \ \ \  m_L=m_R.
\end{equation}
Both parts have the same mass since this is just one particle with two polarization states. At the same time, in the case of a neutral particle the Dirac  bispinor can be split into two real parts  
\begin{equation}
\nu_D=\left(\begin{array}{c} \xi _1\\ \xi_1^* \end{array}\right)+ \left(\begin{array}{c} \xi_2 \\ \xi_2^* \end{array}\right), \ \ \  m_{\xi_1}\neq m_{\xi_2}.
\end{equation}
each of these parts is a Majorana spinor obeying the condition  $\nu_M=\nu_M^*$,
i.e. if neutrino is a Majorana spinor, then it is an antiparticle to itself. These two Majorana spinors can have different masses. Hence, if this possibility is realized in Nature, we have just discovered the light neutrino and the heavy ones can have much bigger masses.

An argument in favour of the Majorana neutrino  is  the smallness of their masses. If one gets them through the usual Brout-Englert-Higgs mechanism, the corresponding Yukawa couplings are extremely small of the order of $10^{-12}$. In the case of the Majorana neutrino one can avoid it using the see-saw mechanism~\cite{seesaw}:  The small masses of light neutrinos appear due to the heaviness of the Majorana mass 
\begin{eqnarray}
&&\hspace{1.3cm} L \ \ \ \ R \nonumber\\
M_\nu&=&\begin{array}{c} L \\ R \end{array}\left( \begin{array}{cc} 0 & m_D \\ m_D & M \end{array}\right), \ \ \  m_1=\frac{m_D^2}{M}, \ m_2= M.
\end{eqnarray}
Thus, the neutrino Yukawa coupling  may have the usual lepton value and the Majorana mass  $M$ might be of the order of the Grand Unification scale. In this case, one also has the maximal mixing in the neutrino sector.

\begin{figure}[ht]
\begin{center}
\leavevmode
\includegraphics[width=0.3\textwidth]{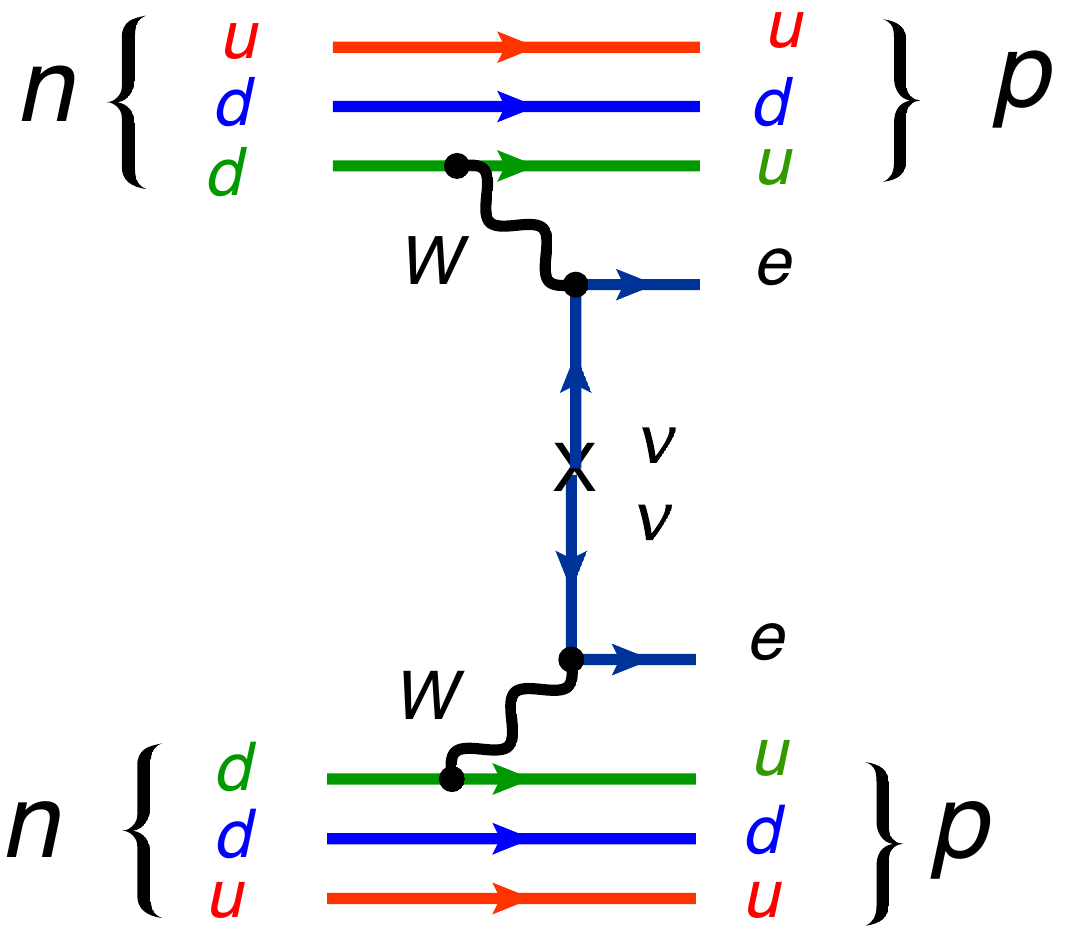}
\includegraphics[width=0.35\textwidth]{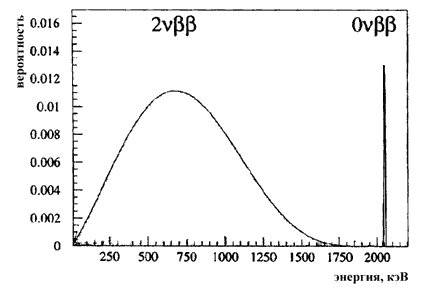}
\includegraphics[width=0.32\textwidth]{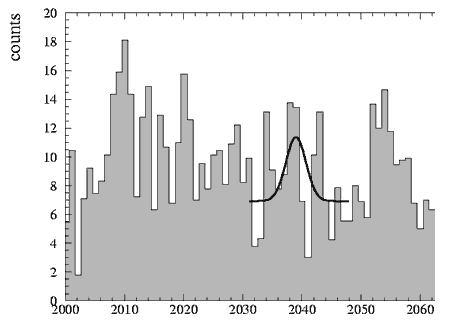}
\end{center}\caption{Neutrinoless double $\beta$- decay (left) and the energy spectrum of electrons in the case of a usual and neutrinoless decay of the isotope $^{76}Ge$ (center).  The experimentally measured spectrum of electrons is shown on the right~\cite{02bb}}
\label{02bb}
\end{figure}

One can find out the nature of the neutrino  studying the double $\beta$-decay. If the neutrinoless double $\beta$-decay is possible, then the neutrino is a Majorana since for the Dirac neutrino it is forbidden. The corresponding Feynman diagram is shown in Fig.\ref{02bb}. It also shows the energy spectrum of electrons in the case of the usual and neutrinoless  $\beta$-decay~\cite{02bb}. 
As one can see, two types of spectrum are easily distinguishable. However, practical observation is rather 
cumbersome.  The histogram shown in Fig.\ref{02bb}, right is the experimentally measured electron spectrum of the double $\beta$-decay. The solid bold line shows the expected position of the maximum in the spectrum of two electrons corresponding to the double neutrinoless $\beta$-decay.

As a result, today there are no clear indications of the existence of the double neutrinoless  $\beta$-decay.  The experiments are carried out on the isotopes  $ ^{48}Ca, ^{76}Ge, ^{82}Se$, $^{130}Te, ^{136}Xe, ^{150}Nd.$  Modern estimates of the lifetime are~\cite{halftime}
\begin{eqnarray*}
T_{1/2}2\nu\beta\beta (^{136}Xe) &\times& 10^{21} \ yr = 2.23\pm 0.017\  stat \pm 0.22 \ sys ,  \\
T_{1/2}0\nu\beta\beta (^{136}Xe) &\times& 10^{25} \ yr > 1.6 \ (90\% \ CL).
\end{eqnarray*} 
Thus, the nature of the neutrino remains an open problem of the SM.

\section{The Flavour sector}

In Fig.\ref{SM},  there are presented  3 generations of matter particles. At the moment, there is no theoretical answer to the question of how many generations exist in Nature.  We have only the experimental facts which can be interpreted  as an indication of the existence of three generations. They assume the presence of the quark-lepton symmetry since refer to the number of light neutrinos and, due to this symmetry, to the number of generations. 

\begin{figure}[ht]
\begin{center}
\leavevmode
\includegraphics[width=0.4\textwidth]{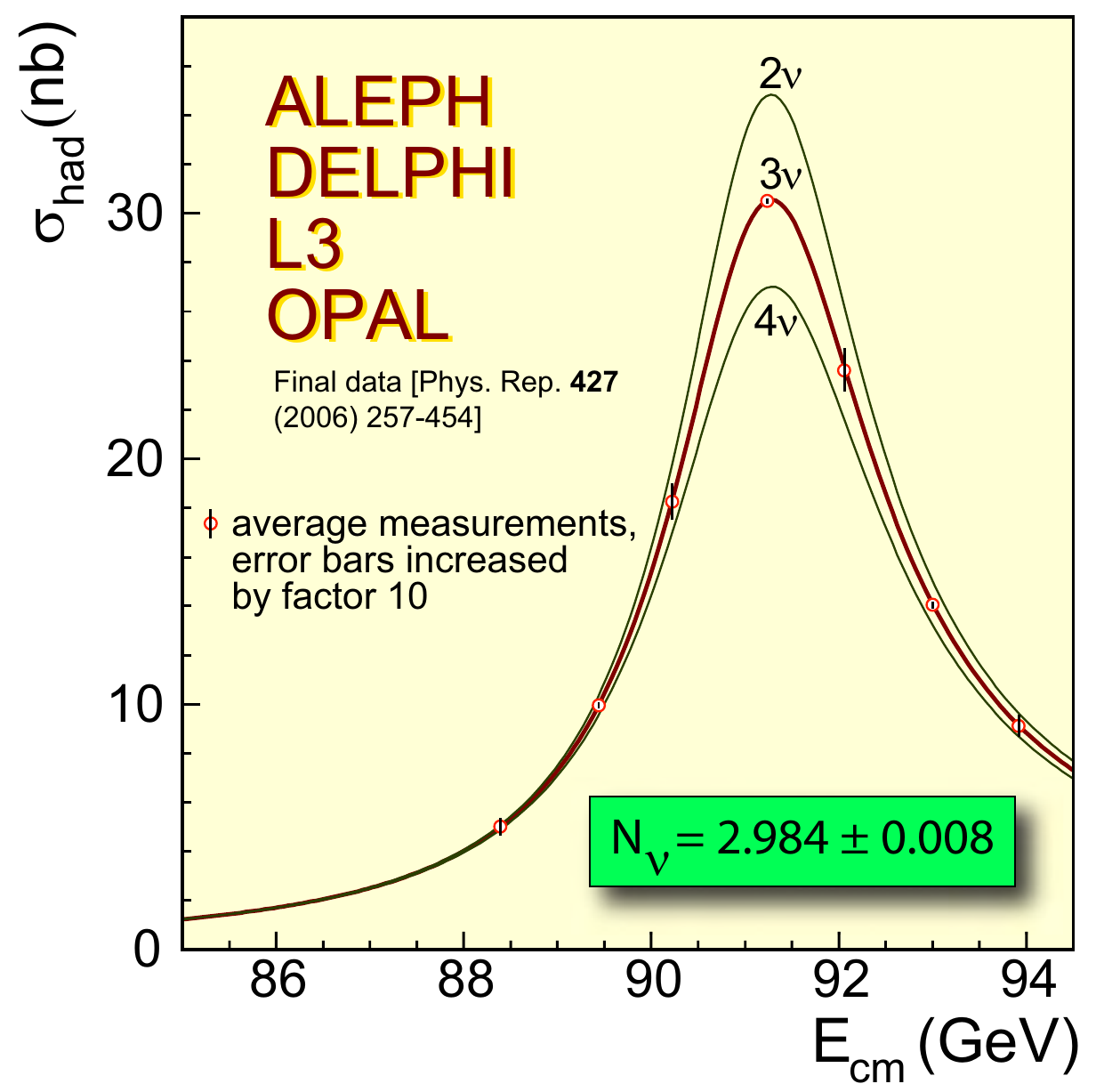}\hspace{1cm}
\includegraphics[width=0.45\textwidth]{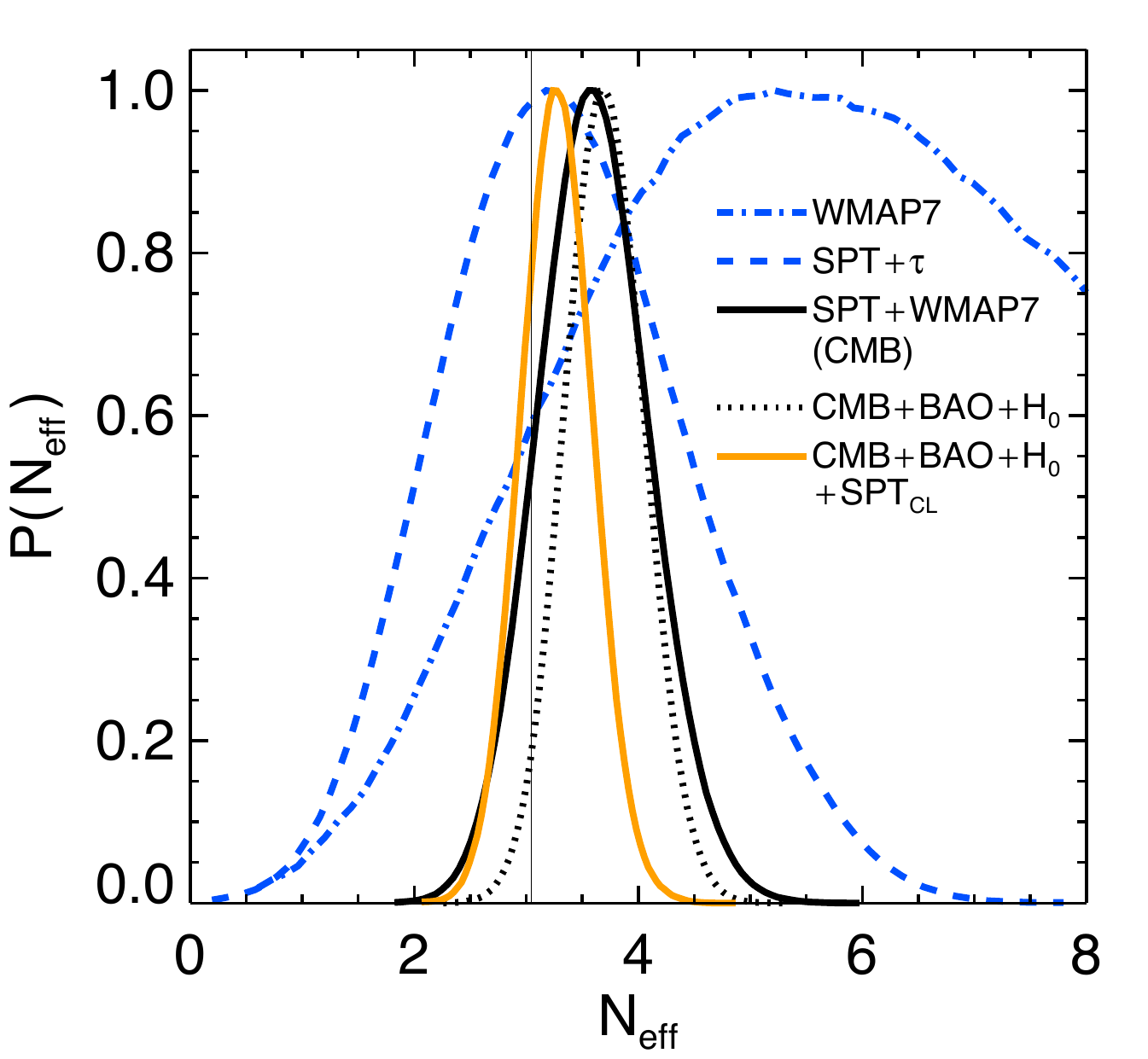}
\end{center}\caption{Experimentally measured profile of  the Z-boson, and the number of light  neutrinos (left) and the fit of the number of light neutrinos   from the temperature fluctuations of CMB (right)}
\label{width}
\end{figure}

The first fact is the measurement at the electron-positron collider LEP of the profile and the width of  the Z-boson. The Z-boson can decay into quarks, leptons and neutrinos with the total mass less than  its own mass and, measuring the width of the Z-boson one can find out the number of light neutrinos.  This is not true for neutrinos with the mass bigger than 45 GeV.  The fit to the data corresponds to the number of neutrinos equal to $N_\nu=2.984\pm0.008$, i.e. 3 (see  Fig.\ref{width} left)~\cite{Z}.

The same conclusion follows from the fit of the spectrum of thermal fluctuations of CMB. The number of light neutrinos as well as the spectra of their masses are reliably defined from the CMB shape  (see Fig.\ref{width} right). The obtained number is:  $N_\nu= < 3.30\pm0.27$~\cite{CMBNeutrino2}, i.e. is also  consistent with 3 but still leave some space for an additional sterile neutrino.

At last, there are complimentary data on precision measurements of the probabilities of rare decays where hypothetical additional heavy quark generations might contribute. According to these data, the fourth generation is excluded at the 90\% confidence level~\cite{4gen}.

A natural question arises:  Why Nature needs 3 copies of quarks and leptons?  All that we see around us is made of protons, neutrons and electrons, i.e. of  $u$ and  $d$ quarks and electrons - particles of the first generation. The particles made of the quarks of the next two generations and heavy leptons, copies of the electron, quickly decay and are observed only in cosmic rays or accelerators. Why do we need them? 

Possibly, the answer to this question is concealed not in the SM but in the properties of the Universe. 
The point is that for the existence of baryon asymmetry of the Universe, which is the necessary condition for the existence of a stable matter, one needs the CP-violation~\cite{Sakharov}. This requirement in its turn is achieved in the SM due to the nonzero phase  in the mixing matrices of quarks and leptons.The nonzero phase appears only when the number of generations  $N_g \geq 3$. The usual parametrization of the mixing matrix in the case of three generations looks like~\cite{PDG}
\begin{equation}
K=\left(\begin{array}{ccc}
V_{ud} & V_{us} & V_{ub} \\
V_{cd} & V_{cs} & V_{cb} \\
V_{td} & V_{ts} & V_{tb} 
\end{array}
\right)=
\left(\begin{array}{ccc}
c_{12}c_{13} & s_{12}c_{13} & s_{13}e^{-i\delta} \\
-s_{12}c_{23}-c_{12}s_{23}s_{13}e^{i\delta}& c_{12}c_{23}-s_{12}s_{23}s_{13}e^{i\delta}& s_{23}c_{13} \\
s_{12}s_{23}-c_{12}c_{23}s_{13}e^{i\delta} & -c_{12}s_{23}-s_{12}c_{23}s_{13}e^{i\delta}& c_{23}c_{13}
\end{array} \right).
\end{equation}
It is this phase  $\delta$ in the quark as well as  lepton mixing matrix is the source of the CP-violation in the SM.

\begin{figure}[ht]
\begin{center}
\leavevmode
\includegraphics[width=0.45\textwidth]{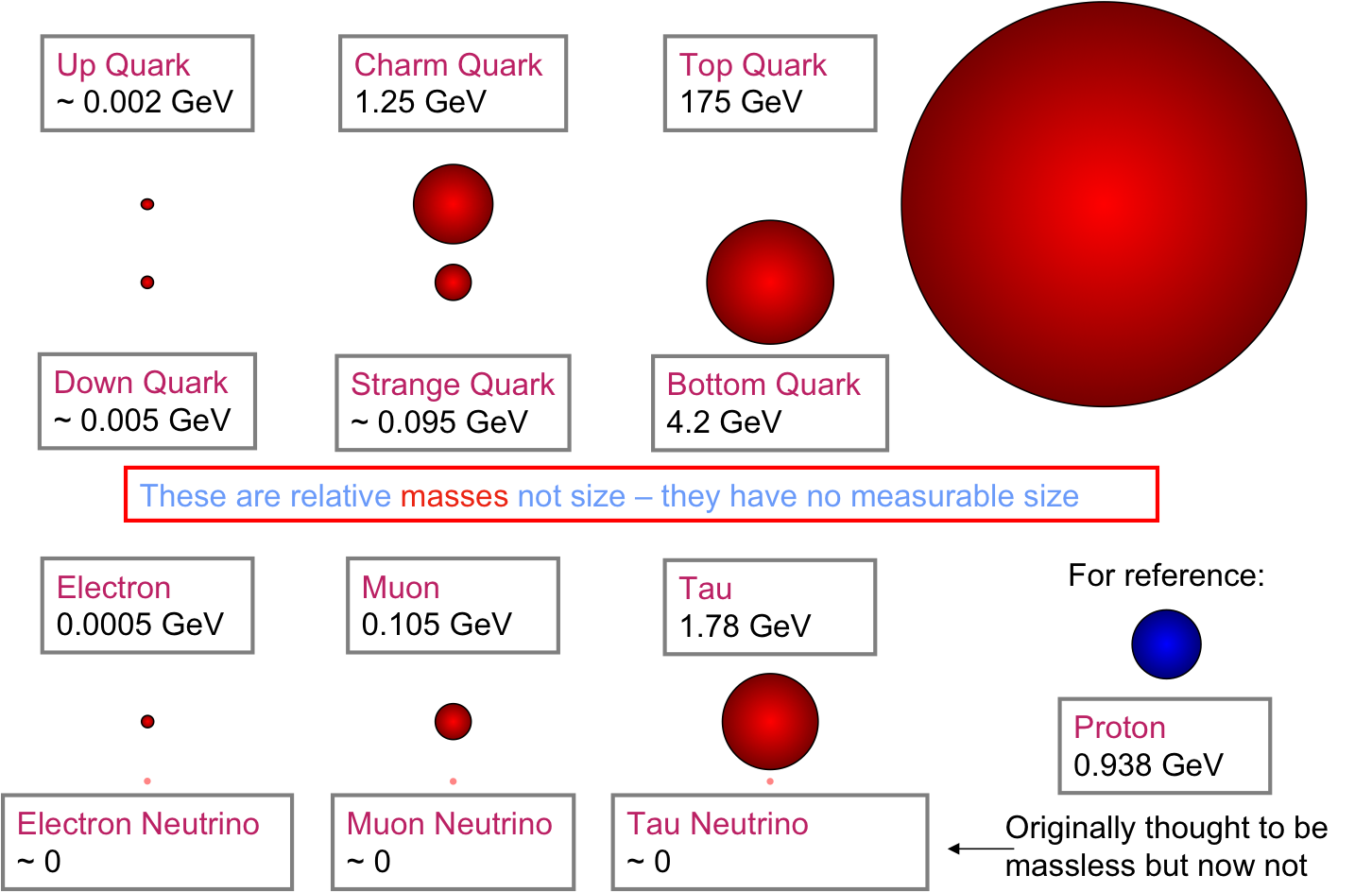}\hspace{1cm}
\includegraphics[width=0.45\textwidth]{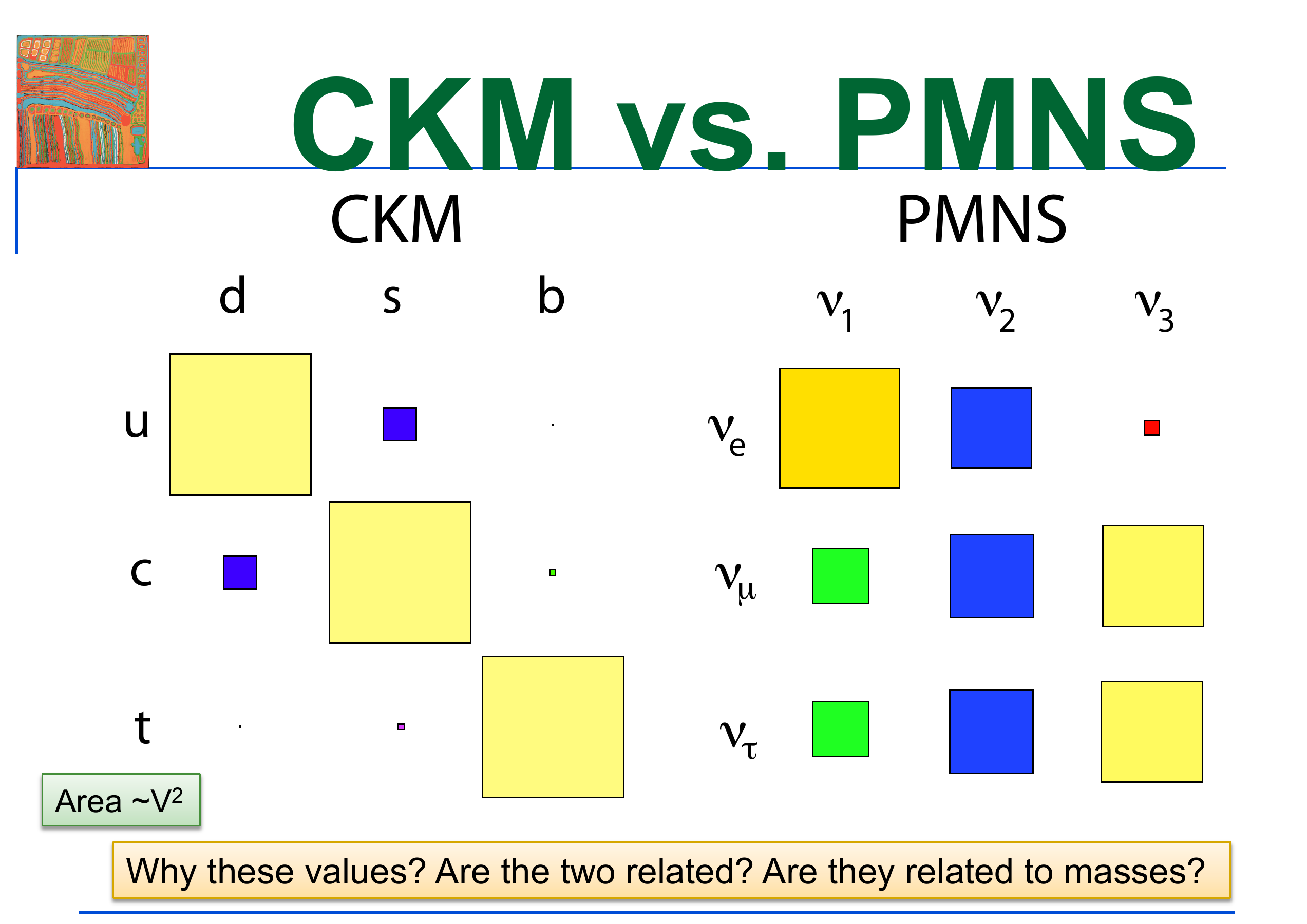}
\end{center}\caption{The mass spectrum of quarks and leptons (left) and the  CKM and the PMNS mixing matrices  (right). The area of the circles and squares is proportional to the numerical values of parameters}
\label{mass}
\end{figure}

The next puzzle of the SM is the mass spectrum of quarks and leptons.  Since the masses of all fundamental particles  in the SM arise from the  vacuum expectation value of a single Higgs field\begin{eqnarray}
m_{quark}=y_{quark} \cdot v, && \nonumber \\
m_{lepton}=y_{lepton} \cdot v,  && \nonumber \\
m_W=g/\sqrt{2} \cdot v,  && \\
m_Z=\sqrt{g^2+g'^2}/\sqrt{2} \cdot v,  && \nonumber \\
m_H=\sqrt{\lambda} \cdot v,  && \nonumber \\
m_{\gamma}=0, && \nonumber \\
m_{gluon}=0, && \nonumber 
\end{eqnarray}
the spectrum of masses is the spectrum of the Yukawa couplings and it is absolutely arbitrary and unclear.Indeed, if one looks at numerical values
 (see  Fig.\ref{mass},  left)~\cite{spectr}, one sees a significant disproportion. The difference in the masses of the first and the third generation achieves three orders of magnitude. The understanding of the mass spectrum remains one of the vital problems of the SM.  

The mixing matrices of quarks ( the Cabibbo-Kobayashi-Maskawa matrix) and leptons (the Pontecorvo-Maki-Nakagawa-Sakato matrix) are equally unclear. If the CKM matrix is almost diagonal, the PMNS matrix is almost uniform (see Fig.\ref{mass}, right)~\cite{matrix}. What explains their big difference?  The phases in both matrices which play the key role in the CP-violation are also unknown. Here possibly lies the answer to the question of the source of the CP-violation: Quark or lepton sector?  The point is that the nonzero phase is usually multiplied by  $sin\theta_{13}$ which is very small in the quark sector but is noticeable in the lepton one.  This may mean that the  {\it baryogenesis}  in fact goes through the  {\it leptogenesis}~\cite{baryogen}. 
This important question  still awaits an answer.
 
\section{Can the SM be valid up to the Planck scale?}

\begin{figure}[htb]
\begin{center}
\leavevmode
\includegraphics[width=0.80\textwidth,height=5.5cm]{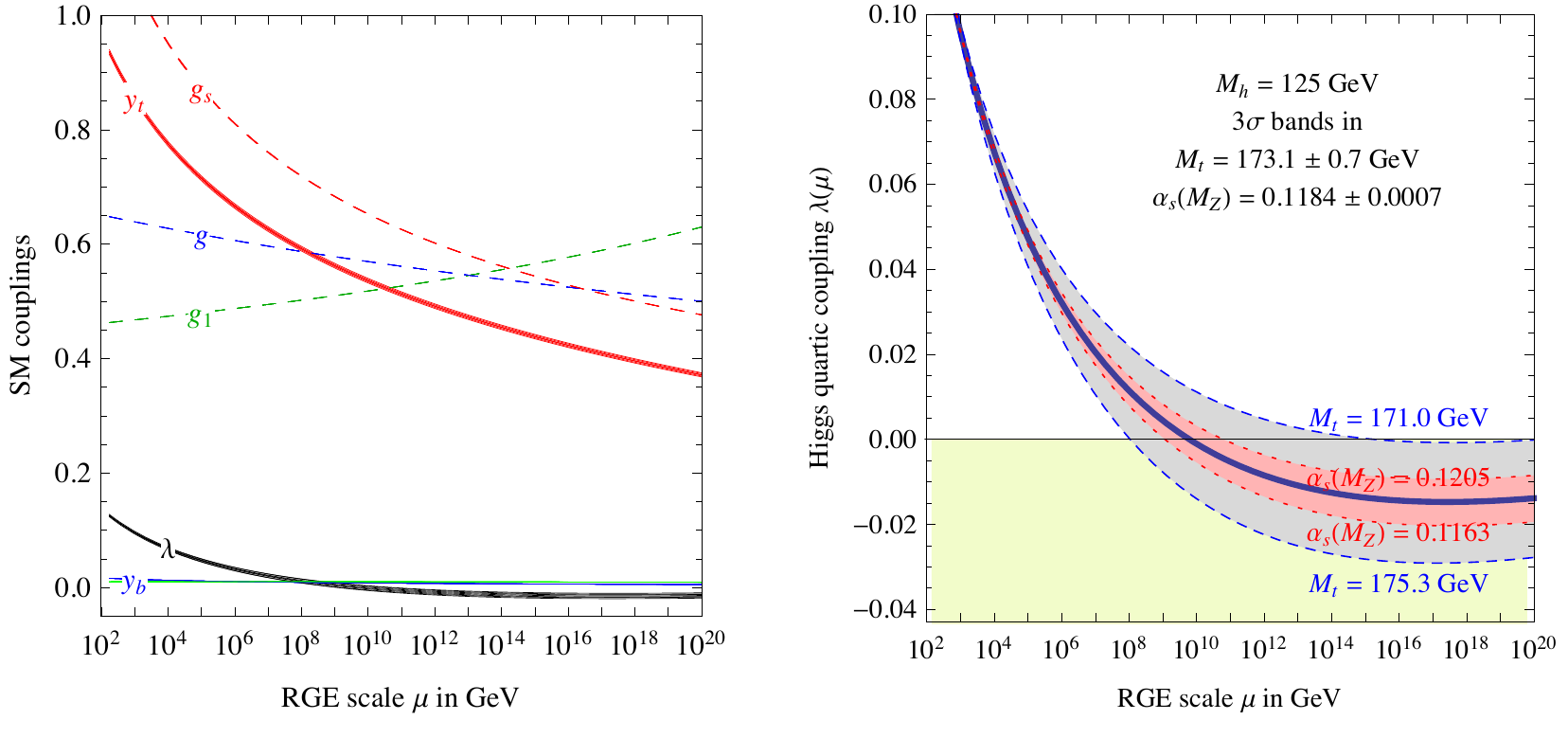}
\end{center}
\caption{The evolution of the gauge couplings $g_1=\sqrt{5/3}g',g_2=g$ and $g_3=g_s$, Yukawa couplings of the third generation $ y_b,y_t$ and the Higgs coupling $\lambda$ (left) and RG evolution of  $\lambda$ 
for the variation of $M_t$ and $\alpha_s$ within  $\pm 3 \sigma$ (right)\cite{Degrassi}}
\label{running}
\end{figure}

The measured mass of the Higgs boson fixes the last unknown parameter of the SM (except, probably, the masses of the Majorana neutrinos) and one can wonder if the SM is valid up to the Planck scale. This means that the parameters of the SM  being continued to the energies of the order of the Planck scale keep finite and do not change the sign so that the theory remains meaningful and has a stable vacuum. To answer this question, consider  the evolution  of the coupling constants of the SM  with energy  from the electroweak scale to the Planck one. The evolution plots for the gauge, Yukawa and Higgs couplings are shown in Fig.\ref{running} (left)~\cite{Degrassi}.  
The zoomed picture of the evolution of the Higgs coupling is shown on the right. One can see that the Higgs coupling crosses zero near the Grand unification scale and the crossing point strongly depends on the values of  $M_t$ and $\alpha_s$.  With account of the two-loop corrections the vacuum stability condition determines the lower bound on the Higgs mass provided the SM is valid up to the Planck scale~\cite{Degrassi}
\begin{equation}
M_h[GeV]>129.4+1.4\left(\frac{M_t[GeV]-173.1}{0.7}\right)-0.5\left(\frac{\alpha_s(M_Z)-0.1184}{0.0007}\right)\pm1.0_{th},
\end{equation}
that gives  $M_h>129.4\pm1.8$  GeV. Thus, the value of 125-126 GeV happens to be slightly lower and the stability condition for the Higgs vacuum is violated at the scale of  $10^{10}-10^{14}$ GeV.

\begin{figure}[htb]
\begin{center}
\leavevmode
\includegraphics[width=0.99\textwidth]{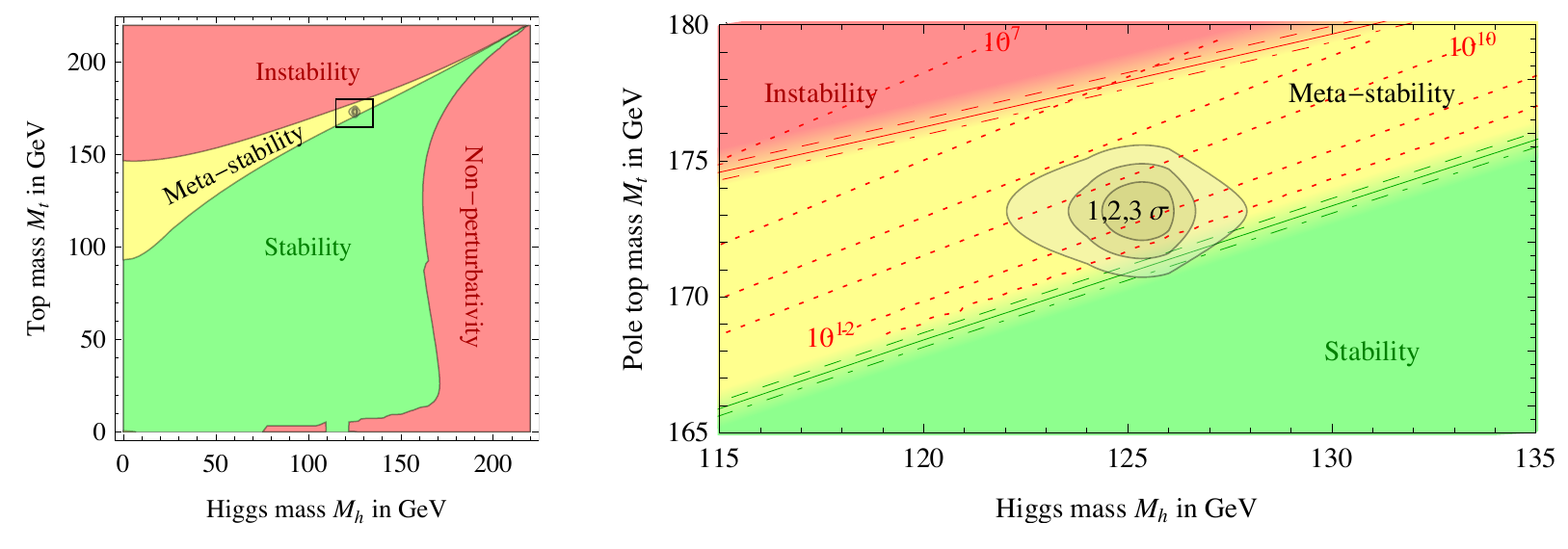}
\end{center}
\caption{The SM vacuum stability, metastability and instability regions in the plane  $M_t-M_H$. The region corresponding to the measured values is zoomed to the right. The dashed lines show the instability scale in GeV}
\label{vacuum}
\end{figure}

The effective potential of the Higgs field  happens to be very sensitive to the values of the Higgs boson and top quark masses. In Fig.\ref{vacuum}~\cite{Degrassi}, it is shown that the measured values of these masses correspond to the point sitting just on the border of a stable and an unstable phases.
Thus, we surprisingly well get into the metastable region on the border of two phases. It should be noted that the presence of a metastable vacuum is not a problem of the SM since the lifetime is rather big. However, the addition of any new particles or new physics at the intermediate scale might essentially change the whole picture. It is interesting that all this happens near the Grand unification scale, which may be occasional but may as well indicate that at this scale some essential changes in the description of Nature take place. 

\section{Dark Matter}

\begin{figure}[ht]
\begin{center}
\leavevmode
\includegraphics[width=0.56\textwidth, height=6.0cm]{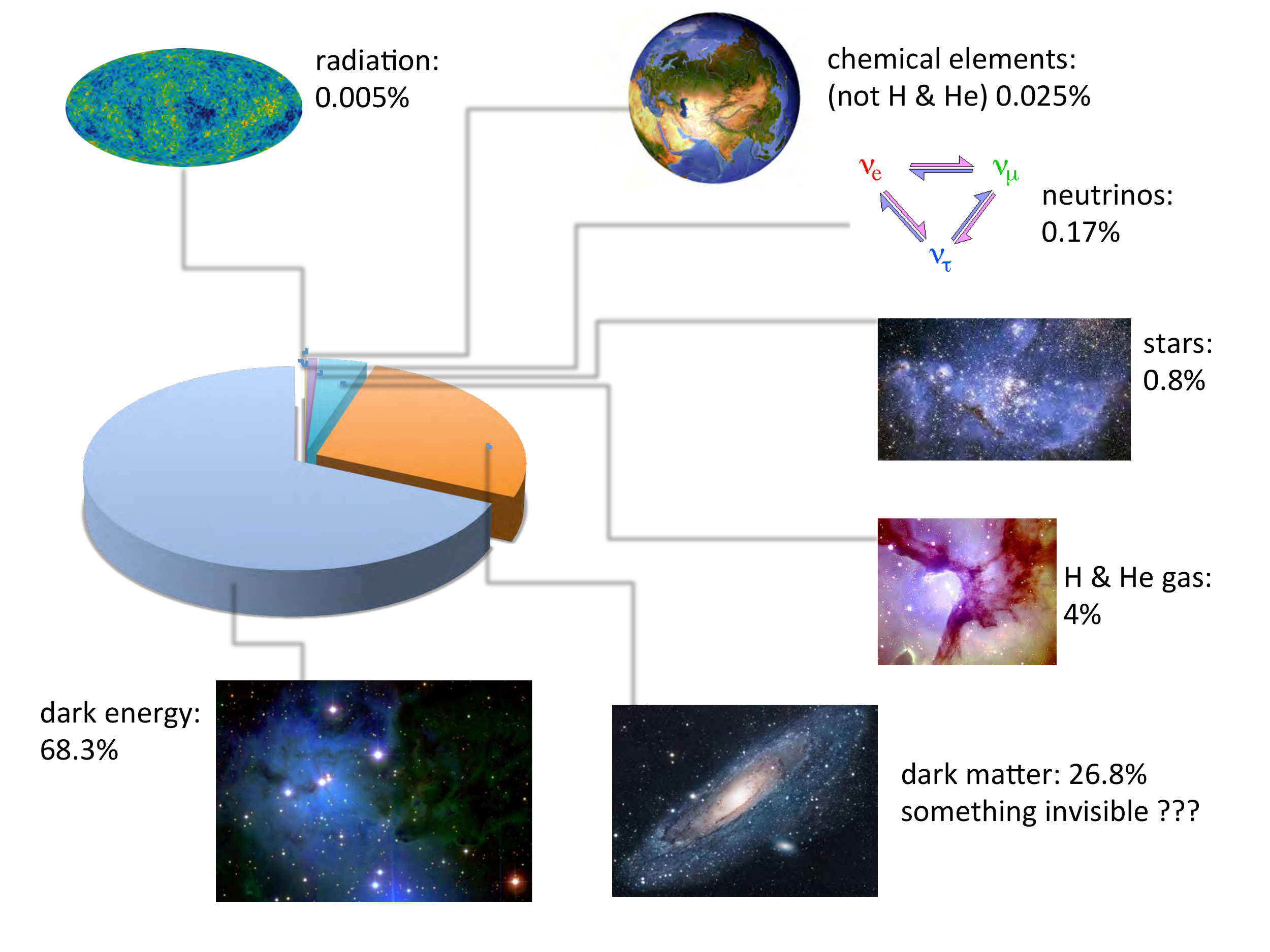}\hspace{0.6cm}
\includegraphics[width=0.38\textwidth, height=4.6cm]{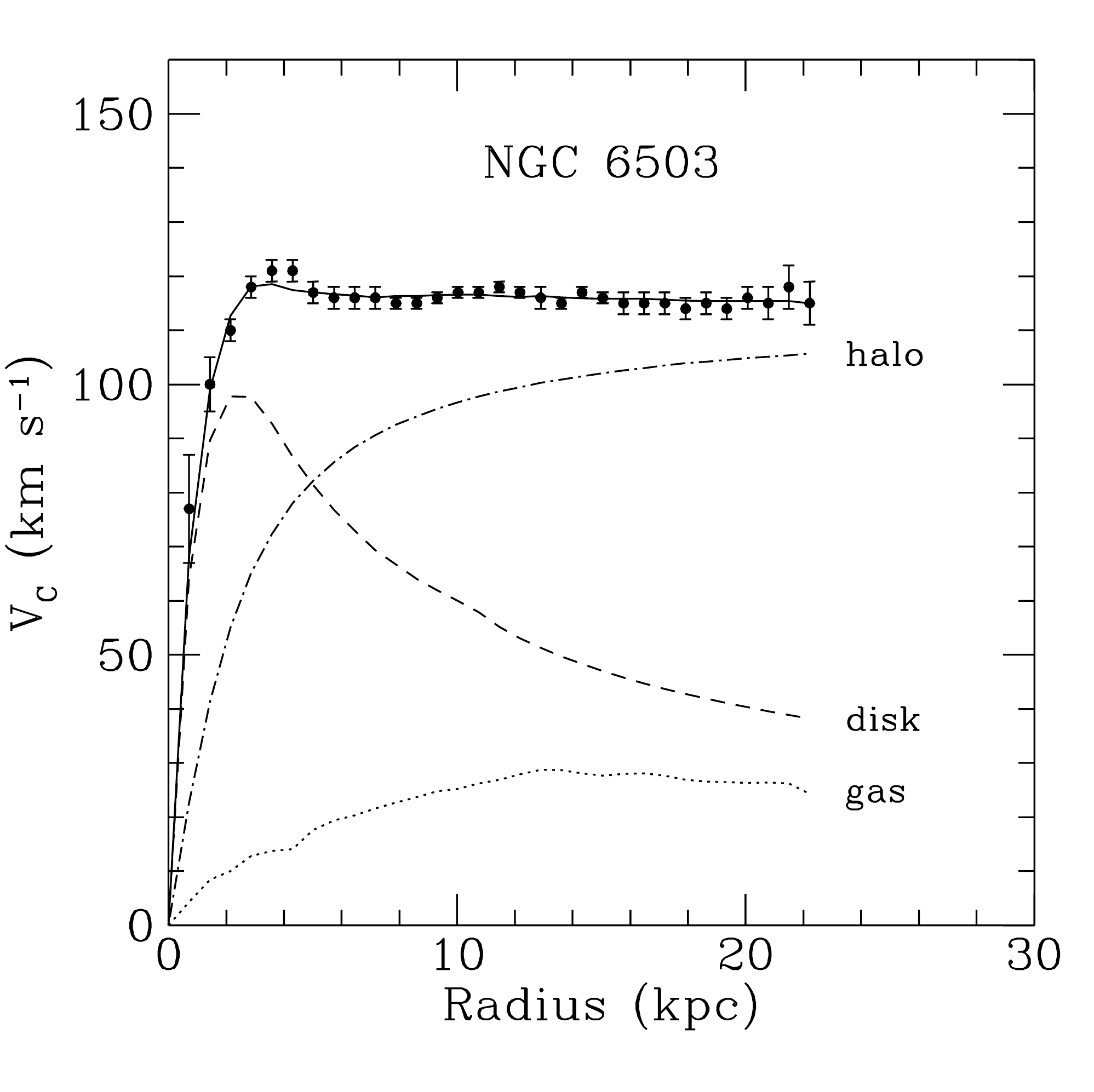}
\end{center}\caption{The energy balance of the Universe (left)~\cite{balance} and rotation curves of stars in the spiral galaxy (right)~\cite{rotation}}
\label{balance}
\end{figure}

The existence of the Dark matter is known since the 30s of the last century. However, the situation has changed when the energy balance of the Universe was obtained and became clear that there is 6 times more  Dark matter  than the ordinary matter (see Fig.\ref{balance}, left)~\cite{WMAP}. The existence of  the Dark matter, which is known so far due to its gravitational influence, is supported by the rotational curves of the stars, galaxies and clusters of galaxies (see Fig.\ref{balance} right), the gravitational lenses, and the large scale structure of the Universe~\cite{DM}. 
Therefore, the question appears: What is the dark matter made of, can it be some non-shining macro objects like the extinct stars, molecular clouds, etc., or these are micro particles? In the last case the Dark matter becomes the object of particle physics. 

According to the last astronomical data, at least in our galaxy, there is no evidence of the existence of macro objects, the so called MACHOs. At the same time, the Dark matter is required for a correct description of the star rotation. Therefore,  the hypothesis of the microscopic nature of the Dark matter is the dominant one. 
In this case, in order to form the large scale structure of the Universe, the Dark matter has to be cold, i.e. nonrelativistic; hence, the DM particles have to be heavy. According to the estimates, their mass has to be above a few dozen of keV~\cite{coldDM}. Besides, the DM particles have to be stable or long-lived to survive since the Big Bang. Thus, one needs a neutral, stable and relatively heavy particle.

If one looks at the SM, the only stable neutral particle is the neutrino. However, if the neutrino is the Dirac particle, its mass is too small to form the Dark matter. Therefore, within the SM the only possibility to describe the Dark matter is the existence of heavy Majorana neutrinos. Otherwise, one needs to assume some new physics beyond the SM. The possible candidates are: 
Neutralino, sneutrino and gravitino in the case of supersymmetric extension of the SM~\cite{susyDM},  and also a new heavy neutrino~\cite{Shap}, a heavy photon, a sterile Higgs boson, etc.~\cite{DMCandidate}. An alternative way to form the Dark matter is there axion field, the hypothetical light strongly interacting particle~\cite{axion}. In this case, the Dark matter differs by its properties. 

The dominant hypothesis is that the Dark matter is made of weakly interacting massive particles - WIMPs. 
This hypothesis is supported by the following fact: The concentration of the Dark matter after the moment when a particle fell down from the thermal equilibrium is given by the Boltzmann equation~\cite{susyDM}
\begin{equation}
\frac{dn_\chi}{dt}+ 3 H n_\chi = - < \sigma v > ( n^2_\chi- n^2_{\chi,eq}),	
\end{equation}
where $H = \dot{R}/ R$  is the Hubble constant, $n_{\chi,eq}$ is the concentration in the equilibrium, and $\sigma$  is the Dark matter annihilation cross-section.The relic density is expressed through the concentration   $n_\chi$  in the following way
\begin{equation}
\Omega_\chi h^2 =\frac{m_\chi n_\chi}{\rho_c}\approx \frac{2\cdot 10^{27}\ cm^3\ sec^{-1}}{<\sigma v>}.
\end{equation}
Having in mind that $\Omega_\chi h^2 \approx 0.113\pm 0.009$ and $v\sim 300$ km/sec,  one gets for the cross-section
\begin{equation}
\sigma\approx 10^{-34}\ cm^2 = 100\ pb,
\end{equation}
that is a typical cross-section for a weakly interacting particle with the mass of the order of the Z-boson mass.

These particles presumably  form an almost spherical galactic halo with the radius a few times bigger than the size of the shining matter. The DM particles cannot leave the halo being gravitationally bounded and cannot stop since they cannot drop down the energy emitting photons like the charged particles.  In the Milky Way, in the region of the Sun the density of the Dark matter should be   $\sim$ 0.3 GeV/sm$^3$ in order to get the observed rotation velocity of the Sun around the center of the galaxy $\sim$ 220 km/sec.

\begin{figure}[h]
\begin{center}%\vspace{-1cm}
\leavevmode
\includegraphics[width=0.6\textwidth]{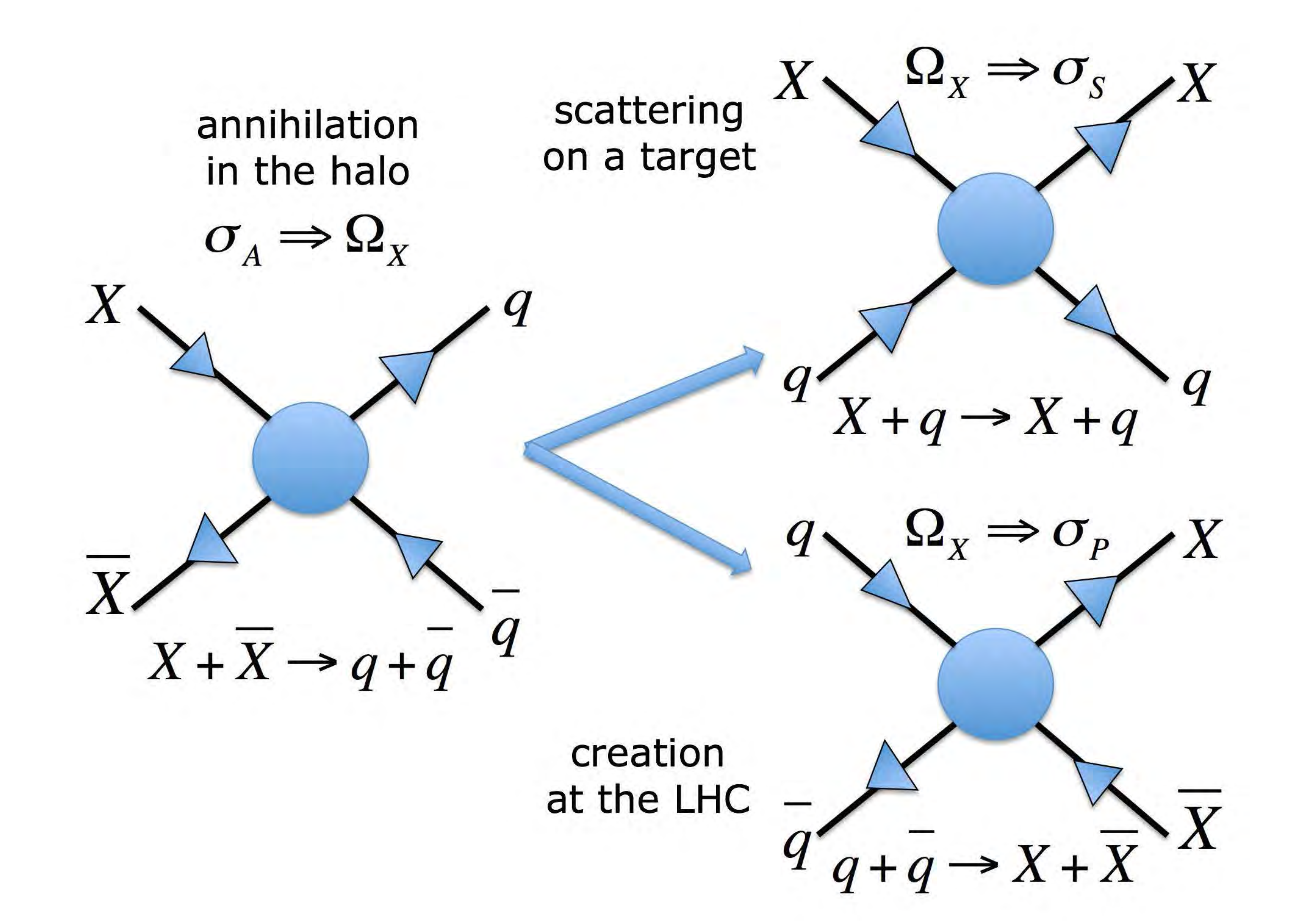}
\end{center}\vspace{-0.5cm}
\caption{The search for the Dark matter}
\label{DMsearch}
\end{figure}

The search for the dark matter particles is based on three reactions the cross-sections of which are related by the crossing symmetry (see Fig.\ref{DMsearch})~\cite{Kolb}. %\phantom{\hspace{4cm}}

This is, first of all, the annihilation of the Dark matter in the galactic halo that leads to the creation of ordinary particles and should appear as the ``knee" in the spectrum of the cosmic rays for diffused gamma rays, antiprotons and positrons.  Secondly, this is the scattering of the DM on the target which should lead to a recoil  of the nucleus of the target when hit by a particle with the mass of the order of the Z-boson mass. And, third, this is a direct creation of the DM particles at the LHC which, due to their neutrality, should manifest themselves in the form of missing energy and transverse momentum.

\begin{figure}[h]
\begin{center}
\leavevmode
\includegraphics[width=0.46\textwidth,height=5.5cm]{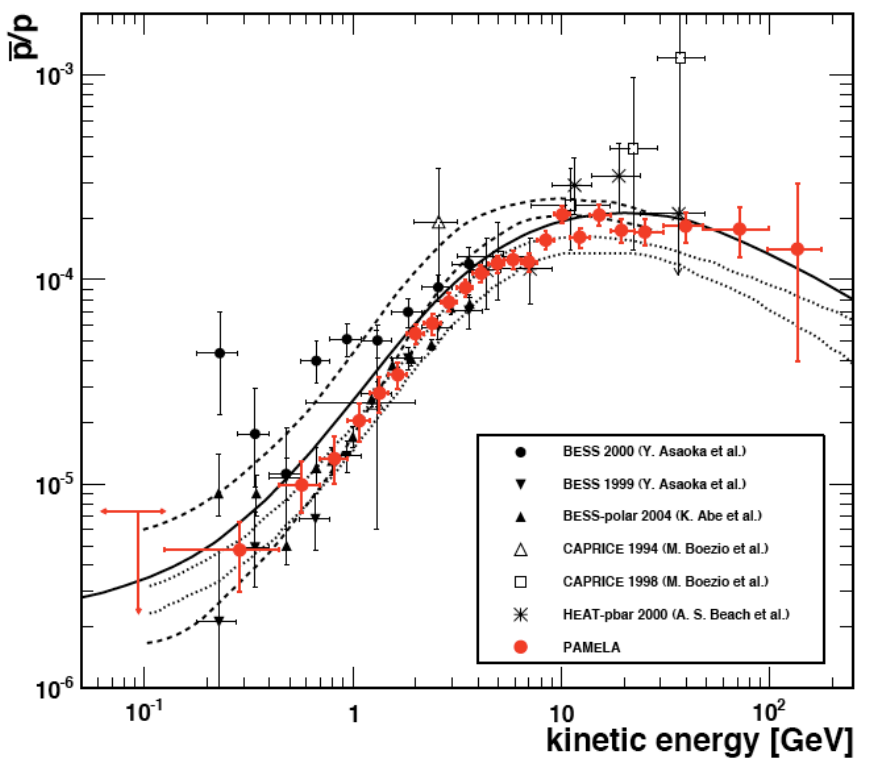}\hspace{0.5cm}
\includegraphics[width=0.44\textwidth,height=5.8cm]{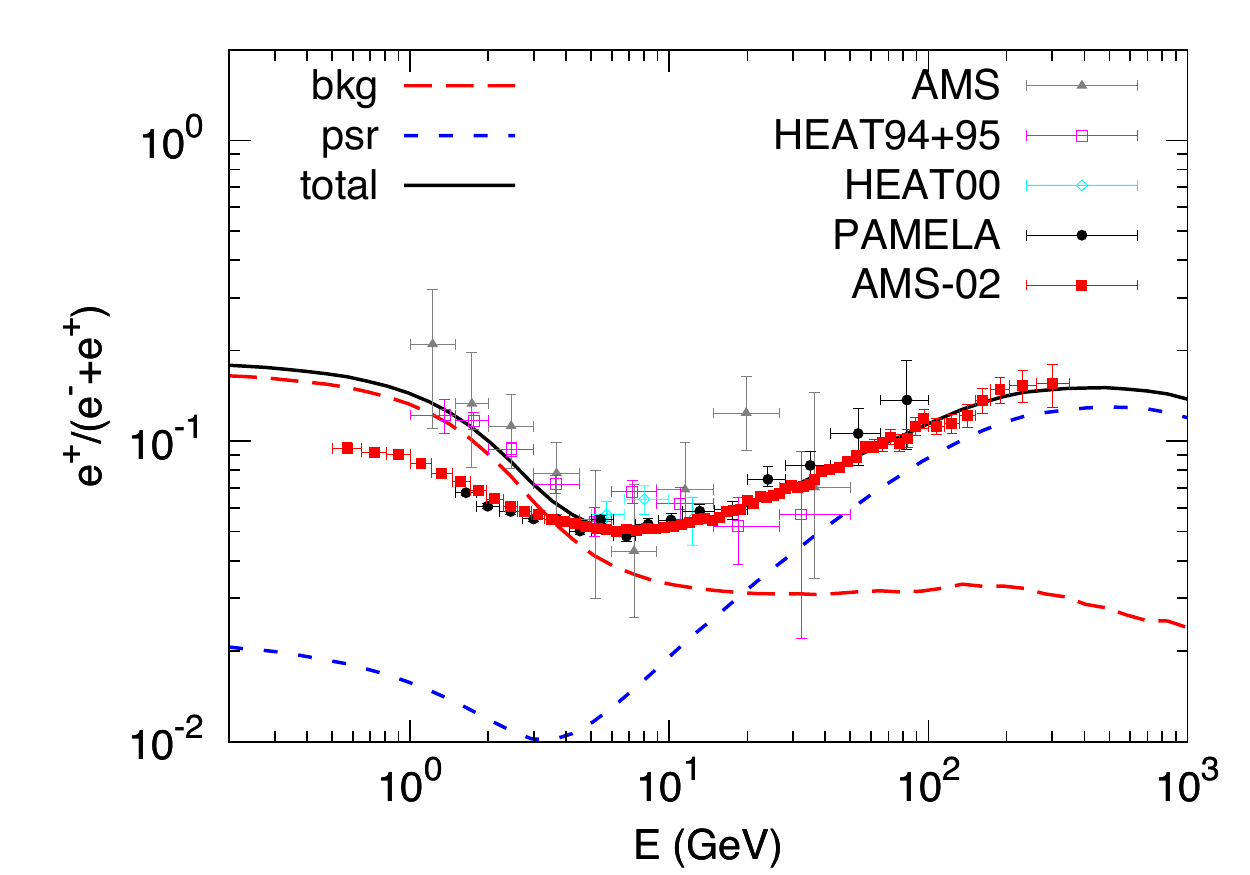}
\end{center}\caption{The search for the Dark matter signal in the spectrum of antiprotons (left) and positrons (right)}
\label{pamela}
\end{figure}

\begin{figure}[htb]
\begin{center}
\leavevmode
\includegraphics[width=0.52\textwidth,height=5.5cm]{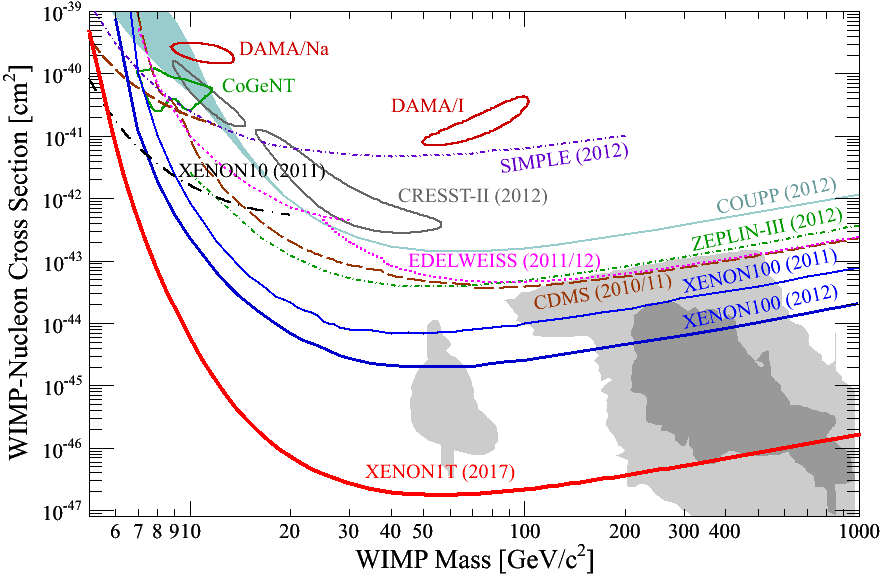}\hspace{1cm}
\includegraphics[width=0.40\textwidth, height=5.5cm]{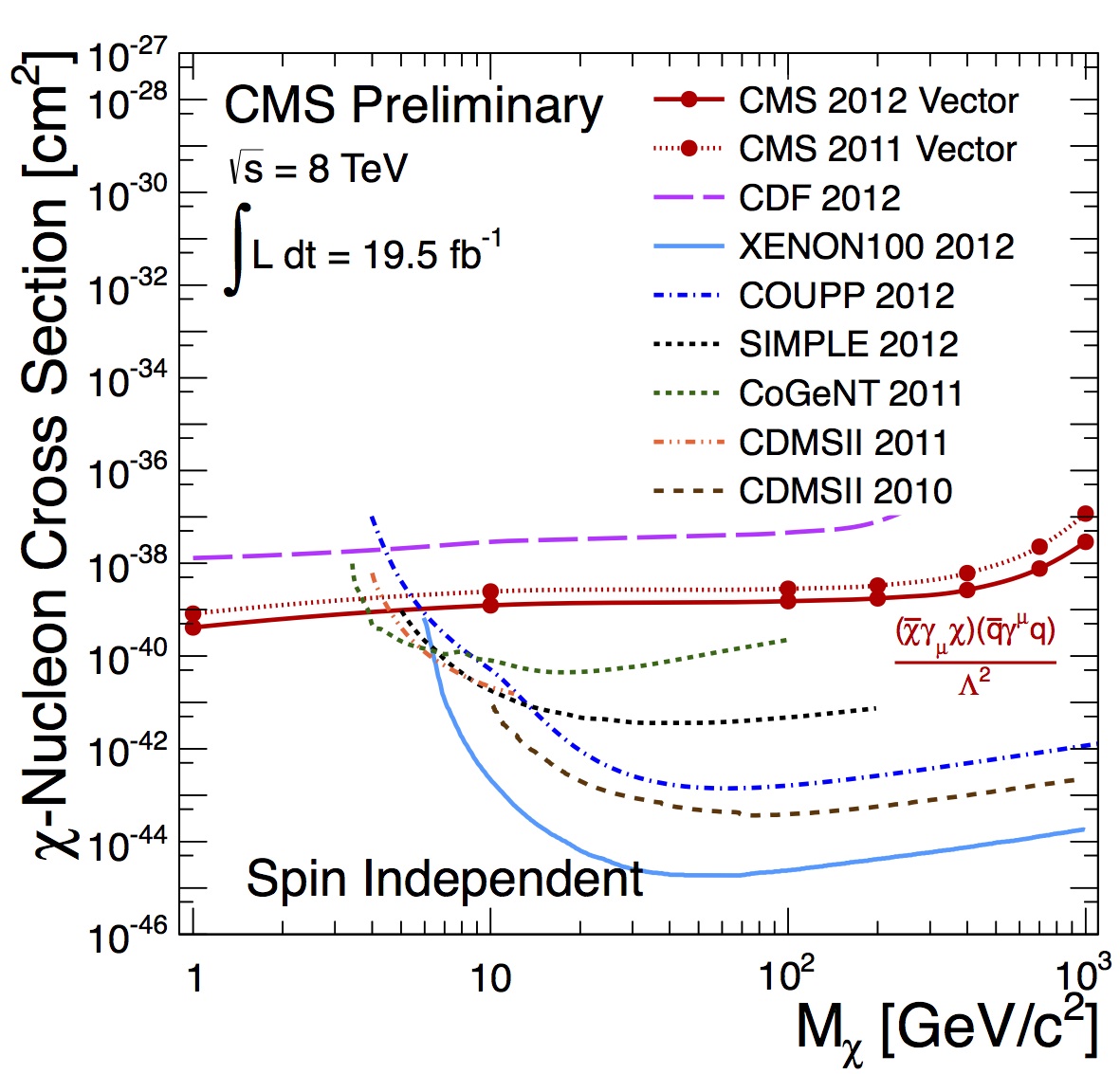}
\end{center}\caption{Direct search for the Dark matter (left) and constraints from the DM creation at  the LHC (right)}
\label{direct}
\end{figure}

In all these directions  there is an intensive search for the signal of the DM. The results of this search for all three cases are shown in Figs.\ref{pamela},\ref{direct}.

As one can see from the cosmic ray data (Fig.\ref{pamela}), in the antiproton sector there is no any statistically significant excess above the background~\cite{Antiproton}. In the positron data there exists some confirmed increase; however,  its origin is usually connected not with the DM annihilation but with the new astronomical source~\cite{positron}. The spectrum of diffused gamma rays like antiprotons is consistent with the background within the uncertainties.

As for the direct detection of the Dark matter, there is  no any positive signal so far. The results of the search are presented in the plane  mass--cross-section. One can see from Fig.\ref{direct}~\cite{direct} that today the cross-sections up to $10^{-45}$ sm$^2$ are reached for the mass near 100 GeV.  In the near future it is planned to advance  two orders of magnitude.

The results of the DM search at the LHC are also shown in the plane mass--cross-section~\cite{LHCDM}. Here the signal of the DM creation is also absent. As it follows from the plot, the achieved bound of possible cross-sections at the LHC  is worse than in  the underground experiments for all mass regions except for the small masses $<10$ GeV where the accelerator is more efficient. Note, however, that the interpretation of the LHC data as the registration of the DM particles is ambiguous and definite conclusions can be made only together with the data from the cosmic rays and direct detection of the scattering of  the DM.

\section{New particles and interactions}

With the achieved TeV energy,  which is one order of magnitude above the electroweak scale, we enter into a new energy region where one can expect the appearance of new particles and  new interactions. However, there is no any guarantee that they exist and it is more intriguing to unveil the mystery.

\begin{figure}[ht]
\begin{center}
\leavevmode
\includegraphics[width=0.75\textwidth]{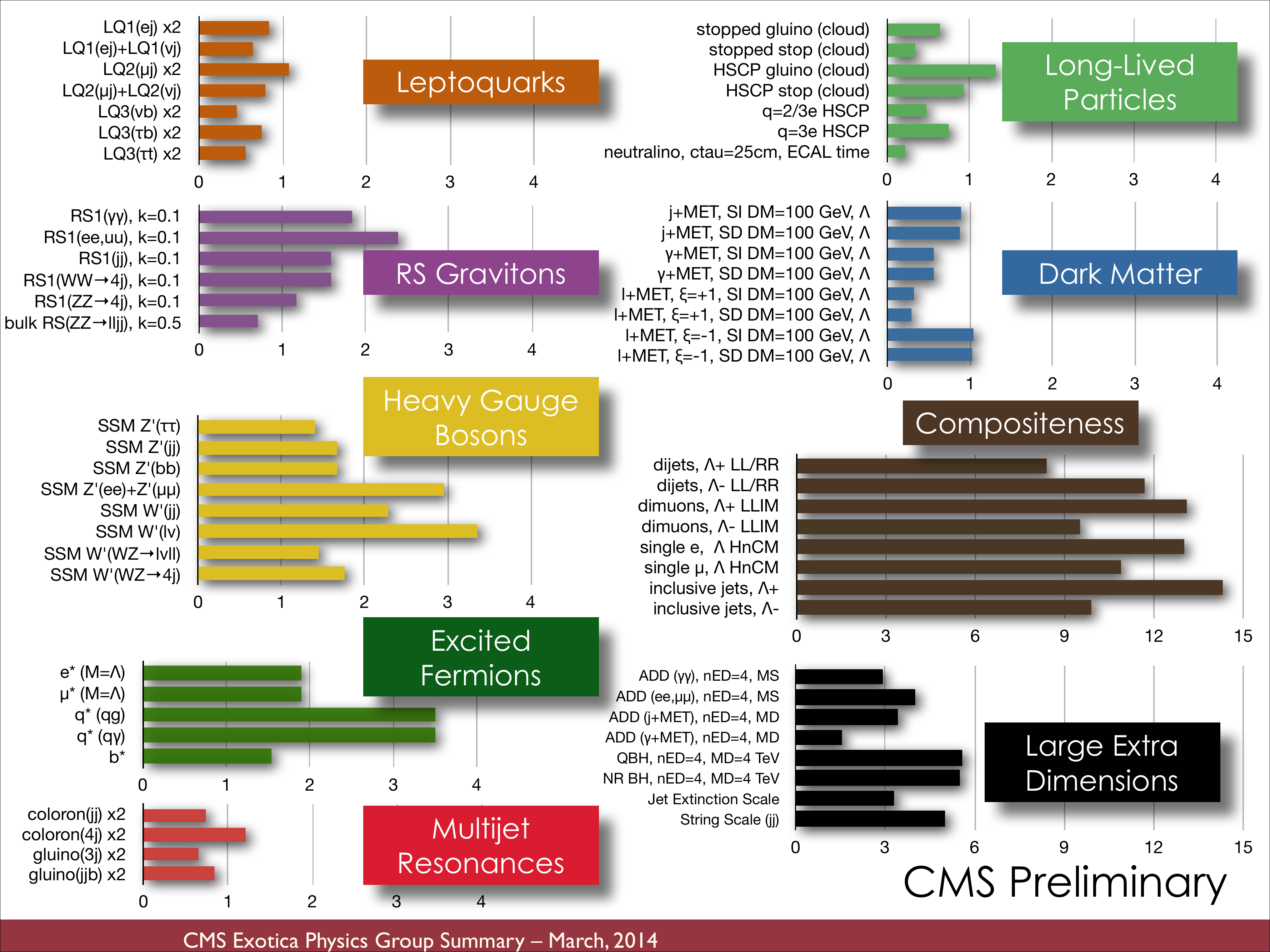}
\end{center}\caption{Search for the manifestation of the new physics at the LHC. 
The constraints in different channels are shown. The numbers are given in TeV}
\label{new}
\end{figure}

There are various suggestions concerning the new physics that may exist at the TeV scale and beyond. They include: Low-energy supersymmetry, extra space-time dimensions, additional gauge symmetries, excited states of quarks, leptons and gauge bosons, lepto\-quarks, exotic hadrons, new heavy generations, long-lived particles, mini black holes, etc. They have different theoretical status and search for the new physics is performed in a wide range. In Fig. \ref{new}, we show modern limits reached in various channels at the accelerator LHC ~\cite{CMS}. So far there are no any signals  of the new physics, but one should remember that we are on the border of  the known reality, on the border of mystery.  Already the very possibility  to look beyond the horizon and  see what is there is incentive!

\begin{figure}[htb]
\begin{center}\vspace{-0.3cm}
\leavevmode
\includegraphics[width=0.65\textwidth]{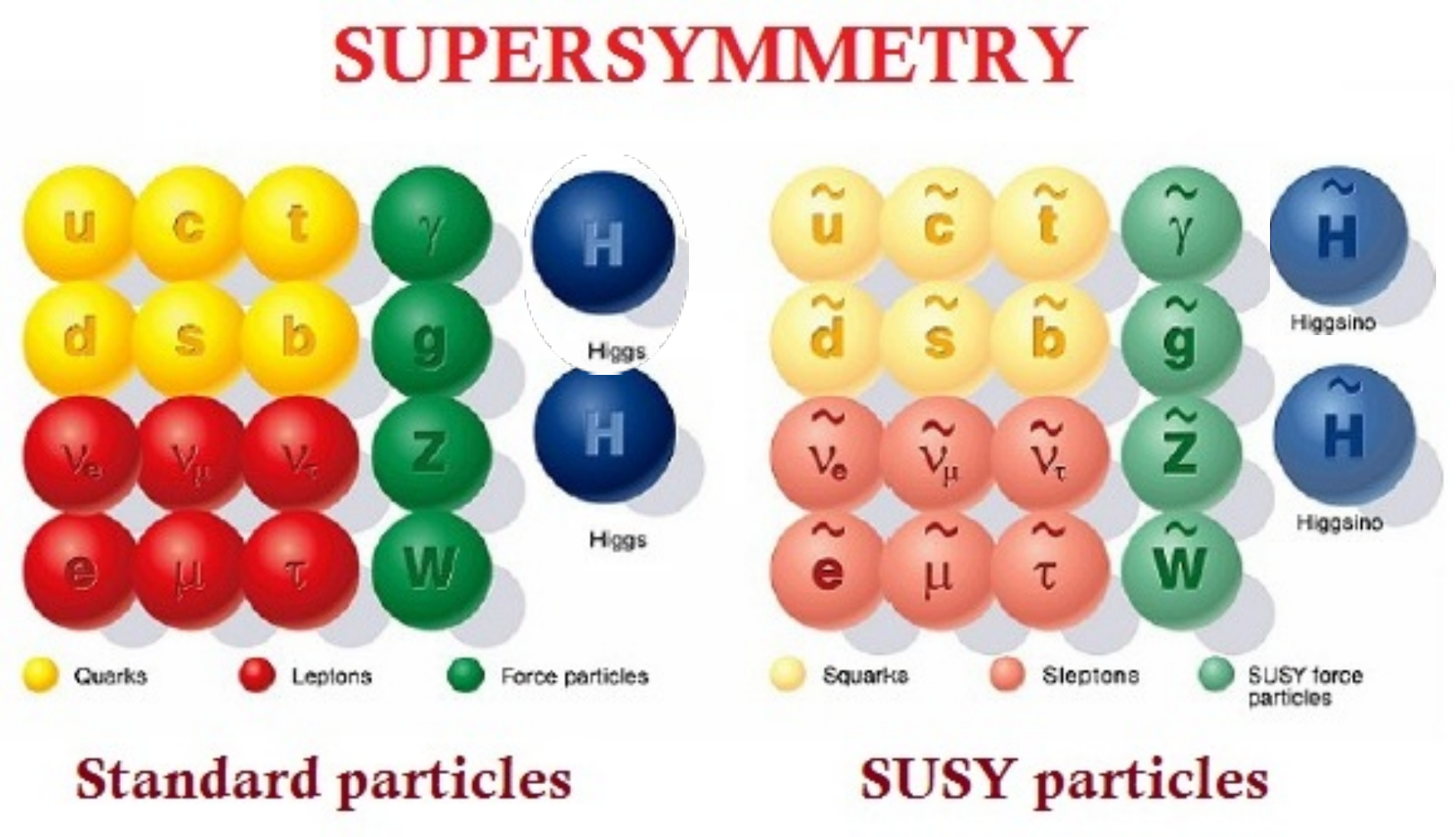}
\end{center}\vspace{-0.3cm}
\caption{The particle content of the minimal supersymmetric model~\cite{susypic}}
\label{susy}
\end{figure}

The most discussed and  most expected new physics is the low-energy supersym\-metry~\cite{MSSM}. There are several reasons why supersymmetry attracts attention of theorists and experimentalists. However, the main reason, from our point of view, is that supersymmetry is a dream of a unified theory of  all  the known interactions  including gravity. The specific feature of supersymmetric theories is the doubling of particles: Each particle of the SM has its own partner, called superpartner, with the same quantum numbers but with spin different by 1/2 (see Fig.\ref{susy}).  The MSSM contains also two Higgs doublets and the corresponding higgsinos.

\clearpage

Let us remind what is  remarkable in TeV scale supersymmetry and what is remarkable in supersymmetry in general.

\noindent Supersymmetry at TeV scale: 
\begin{itemize}
\item Leads to unification of the gauge coupling constants (GUT);
\item Solves the hierarchy problem in the Higgs sector;
\item Provides the electroweak symmetry breaking.
\end{itemize}
Supersymmetry in particle physics:
\begin{itemize}
\item enables inclusion of  gravity in the unified theory;
%
%$\bullet$ 
\item 
provides the existence of the Dark matter;
%
%$\bullet$
\item  stabilizes the string theory as a basis of a unified scheme.
%$\bullet$ 
\end{itemize}

As a rule, the predictions of the superpartner mass spectrum is based on the so called naturalness assuming the natural hierarchy of masses of strongly and weakly interacting particles. Note, however, that all predictions are to a great extent model dependent though this is also true for the analysis of experimental data.  \phantom{\hspace{2cm}}
\begin{wrapfigure}{l}{0.55\linewidth} 
\begin{center}
\leavevmode\vspace{-0.5cm}

\includegraphics[width=0.5\textwidth]{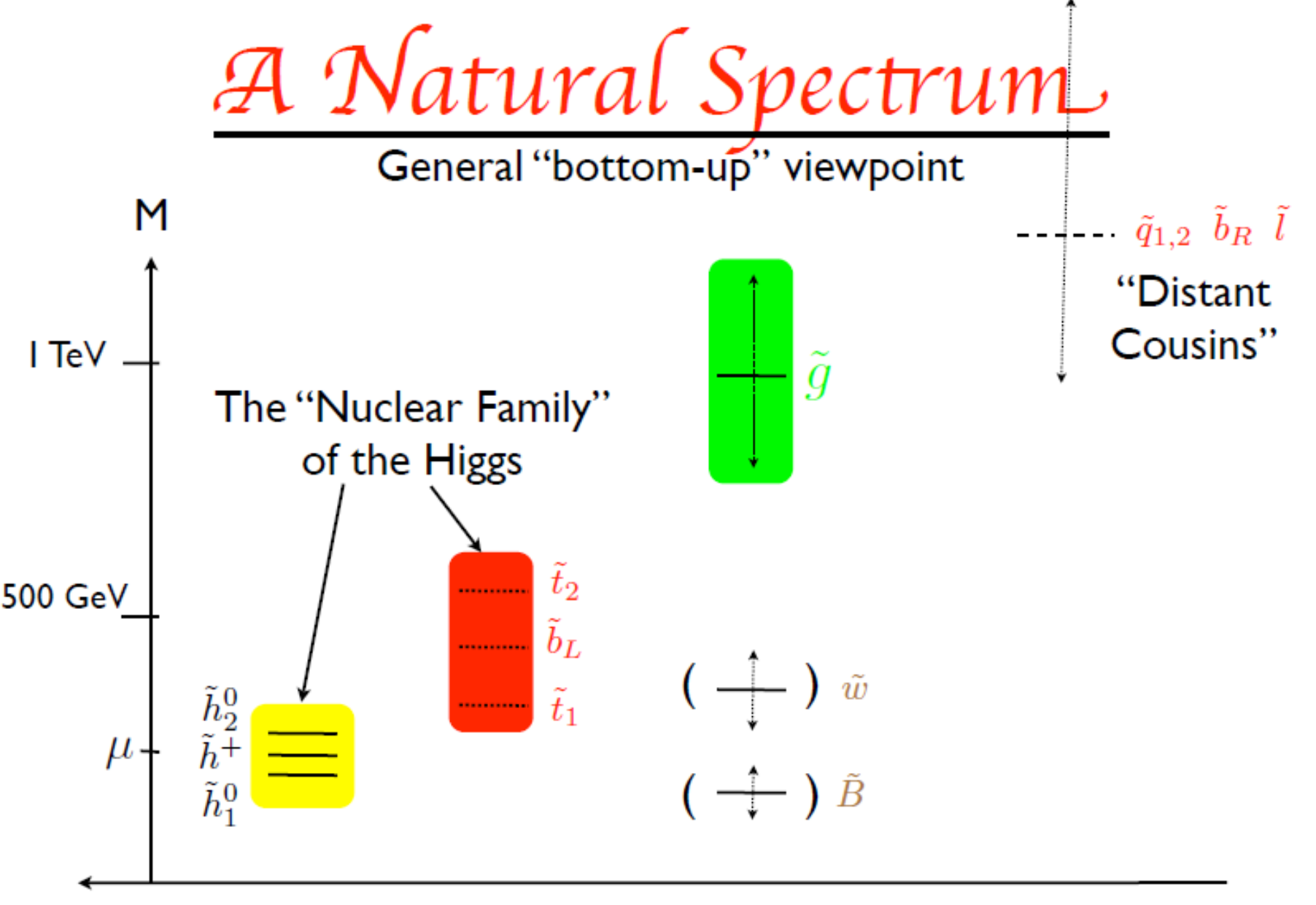}
\end{center}\vspace{-0.1cm}
\caption{The typical ``natural"  mass spectrum of superpartners~\cite{natur}}
\label{natural}
\end{wrapfigure}
The weakest point of modern supersym\-metric extensions of the SM is the problem of supersymmetry breaking. The scheme accepted today, based on a hidden sector, contains large arbitrariness and strongly depends on a particular mechanism. The most natural and developed method of SUSY breaking is the mechanism of spontaneous breaking in the gravity sector with a subsequent transfer of breaking into the visible sector due to the gravitational interaction. In this case, the ``natural" scenario is realized.

\begin{figure}[htb]
\begin{center}
\leavevmode
\includegraphics[width=0.8\textwidth]{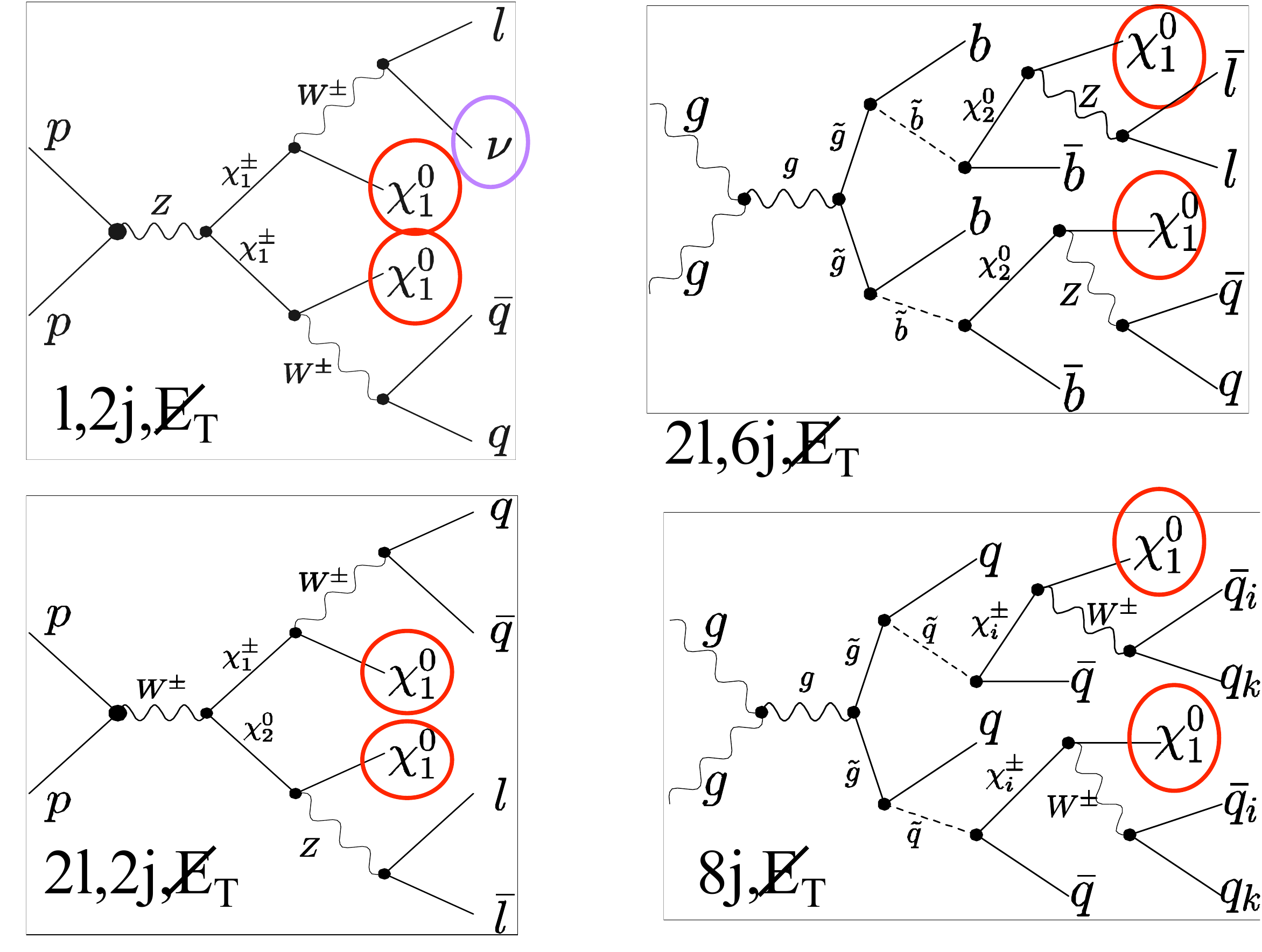}
\end{center}
\caption{Creation of superpartners in weak (left) and strong (right) interactions. The expected final states are also shown}
\label{sprod}
\end{figure}

Under the assumption that super\-symmetry exists at the TeV scale the superpartners of ordinary particles have to be produced at the LHC. The typical processes of creation of superpartners in strong and weak interaction are shown in Fig.\ref{sprod}~\cite{MSSM1}.
The typical signature of supersymmetry is the presence of missing energy and missing transverse momentum carried away by the lightest supersymmetric particle  $\chi^0_1$ which is neutral and stable.

Search for supersymmetry is performed in direct experiments with the creation of superpartners at colliders as well as in precision measurements of low-energy processes where supersymmetry might have indirect manifestation and also in astrophysical and under\-ground experiments.

\begin{figure}[ht]
\begin{center}
\leavevmode \vspace{-1cm}
\includegraphics[width=0.4\textwidth]{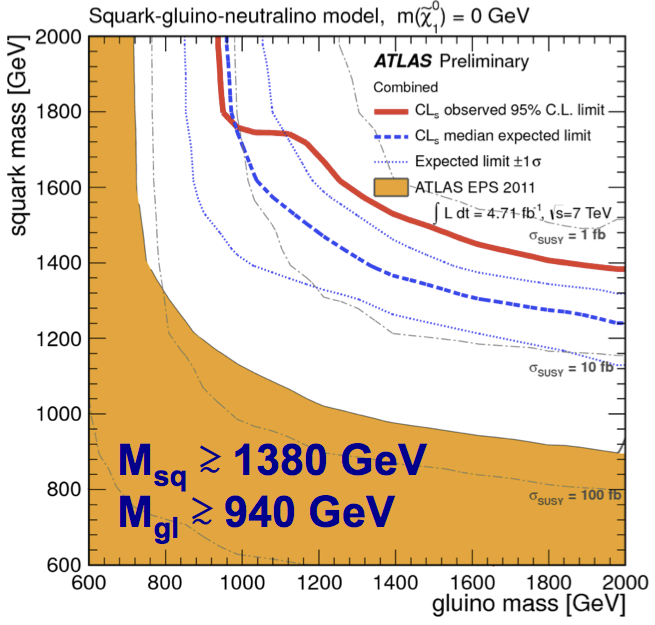}
\includegraphics[width=0.4\textwidth]{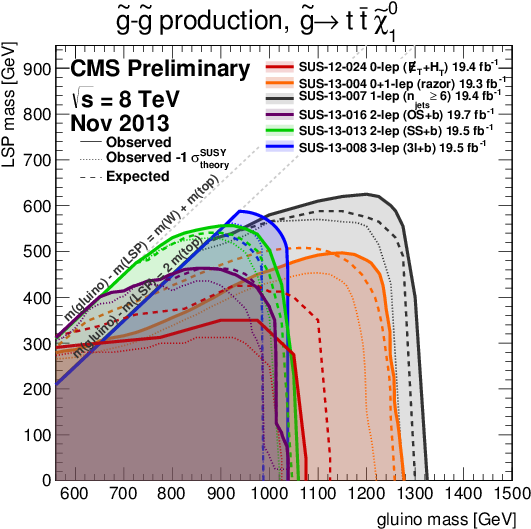}\vspace{1.2cm}
\includegraphics[width=0.4\textwidth]{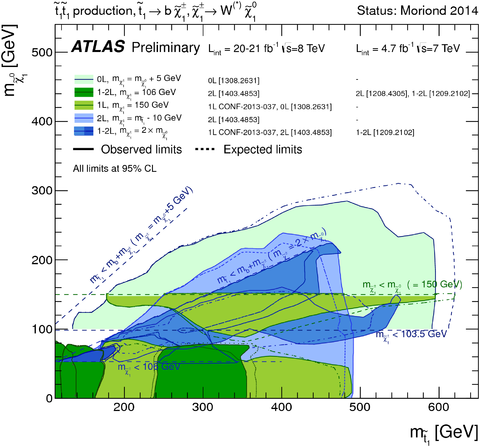}
\includegraphics[width=0.4\textwidth]{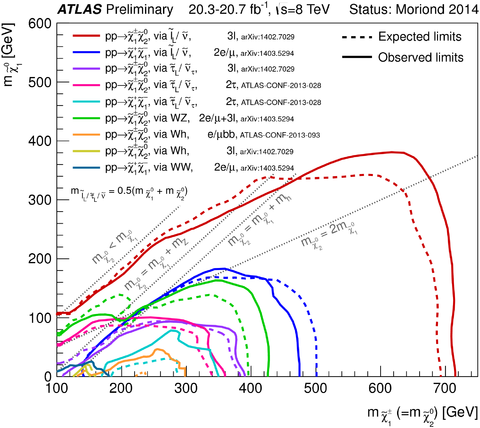}
\end{center}
\caption{Search for supersymmetry at the LHC. There are shown the mass limits for the strongly interacting (up) and weakly interacting (down) particles~\cite{susylhc}}
\label{LHC}
\end{figure}

So far the creation of superpartners at the LHC is not found, there are only limits on the masses of the hypothetical new particles. As one can see from Fig.\ref{LHC}, the progress achieved during one year of the LHC run is rather remarkable. The boundary of possible values of masses of the scalar quarks and gluino have reached approximately 1500 and 1000 GeV, respectively.  For the stop quarks it is almost two times lower. This is because the created squark always decays into the corresponding quark and in the case of the top quark, due to its heaviness, the phase space decreases and so does the resulting cross-section. For the lightest neutralino the mass boundary varies between 100 and 400 GeV depending on the values of the other masses. The constraints on the masses of charged weakly interacting particles  almost two times higher than those for the neutral ones but depend on the decay mode. Let us stress once more that the obtained mass limits depend on the assumed decay modes which in their turn depend on the mass spectrum  of superpartners, which is unknown.  The presented constraints refer to the natural scenario. 

Still, the enormous progress reached by the LHC is slightly disappointing.  The natural question arises: Are we looking in the right direction? Or maybe we have not yet reached the needed mass interval? The answers to these questions can be obtained at the next runs of the accelerator. For the doubled energy the cross-sections of the particle production  with the masses around 1 TeV rise almost by an order of magnitude, and one might expect much higher statistics.

The conclusions that can be made today are~\cite{Kazakov}:
\begin{itemize}
\item So far we do not see supersymmetry,
 \item The obtained constraints are model dependent,
 \item The model contains many parameters and there is still plenty space for supersymmetry,
\item  It is possible that another scheme of SUSY breaking is realized,
\item  The run of the accelerator at maximal energy of 14 TeC in 2015-2016 will be crucial for the discovery of low-energy supersymmetry.
\end{itemize}

\section{Conclusion. Forward into the future}

Thus, the Standard Model of fundamental interaction created, calculated out and experi\-men\-tally tested  during the last 50 years and  triumphantly completed with the discovery of the Higgs boson, is still hiding many mysteries and unresolved problems. Their solution requires big efforts for many years, and possibly during their exploration  many new particles and new interactions will be discovered that will lead to the extension of the SM.

The nearest tasks (at the LHC) are~\cite{Zwirner}:
\begin{itemize}
\item The study of the properties of the new scalar particle with maximal possible precision,
\item The search for any possible deviations from the SM indicating the existence of  new physics,
\item Direct search for new physics at TeV scale.
\end{itemize}
The fulfillment of this program might require the construction of a new electron-positron collider in addition to the existing hadron collider.
 
One should not forget the problem of flavor. The flavor sector of the SM is empirical and has not got proper theoretical understanding so far.

This program has to include also nonaccelerator experiments  investigating neutrino physics and search for the Dark matter, astrophysical experiments unravelling the properties of the Universe as well as the program of studying the structure of hadron matter in collisions of heavy ions. 

We live in exciting time and have a chance to unveil the mystery!

\section*{Acknowledgments}
The author is grateful to V.A.Rubakov for the invitation to give the present  talk. Thanks are also to S.H.Tanyildizi for his help in preparing the manuscript. The financial support from the RFBR, grant \# 14-02-00494 is acknowledged.

\end{document}